\documentclass[prb,twocolumn,letterpaper,superscriptaddress]{revtex4-1}
\usepackage[space]{grffile}
\usepackage{bm,amsmath,color}
\usepackage{multirow,tabularx}
\usepackage{longtable}
\usepackage{stix,pifont}
\usepackage{xcolor}
\usepackage{float}  
\usepackage{changes}
\graphicspath{{./figs/png/}}
\usepackage[cal=cm]{mathalfa}
\def\version{2.9}

\def\Vbckgnd{\ensuremath \overline{\mathscr{V}}_2}

\begin{document}

\title{Nonlocal magnon transconductance in extended magnetic insulating films.\\ I: spin diode effect.} 

\author{R. Kohno}
\affiliation{Université Grenoble Alpes, CEA, CNRS, Grenoble INP, Spintec, 38054 Grenoble, France}

\author{K. An}
\affiliation{Université Grenoble Alpes, CEA, CNRS, Grenoble INP, Spintec, 38054 Grenoble, France}

\author{E. Clot}
\affiliation{Université Grenoble Alpes, CEA, CNRS, Grenoble INP, Spintec, 38054 Grenoble, France}

\author{V. V. Naletov} 
\affiliation{Université Grenoble Alpes, CEA, CNRS, Grenoble INP, Spintec, 38054 Grenoble, France}

\author{N. Thiery}
\affiliation{Université Grenoble Alpes, CEA, CNRS, Grenoble INP, Spintec, 38054 Grenoble, France}

\author{L. Vila}
\affiliation{Université Grenoble Alpes, CEA, CNRS, Grenoble INP, Spintec, 38054 Grenoble, France}

\author{R. Schlitz}
\affiliation{Department of Materials, ETH Zürich, 8093 Zürich, Switzerland}

\author{N. Beaulieu} 
\affiliation{LabSTICC, CNRS, Universit\'e de Bretagne Occidentale,
  29238 Brest, France}

\author{J. Ben Youssef} 
\affiliation{LabSTICC, CNRS, Universit\'e de Bretagne Occidentale,
  29238 Brest, France}
  
\author{A. Anane} 
\affiliation{Unit\'e Mixte de Physique CNRS, Thales, 
    Univ. Paris-Sud, Universit\'e Paris Saclay, 91767 Palaiseau, France}

\author{V. Cros} 
\affiliation{Unit\'e Mixte de Physique CNRS, Thales, 
    Univ. Paris-Sud, Universit\'e Paris Saclay, 91767 Palaiseau, France}

\author{H. Merbouche} 
\affiliation{Unit\'e Mixte de Physique CNRS, Thales, 
    Univ. Paris-Sud, Universit\'e Paris Saclay, 91767 Palaiseau, France}

\author{T. Hauet} 
\affiliation{Université de Lorraine, CNRS Institut Jean Lamour, 
    54000 Nancy, France}

\author{V. E. Demidov}
\affiliation{Department of Physics, University of Muenster, 48149 Muenster, Germany}

\author{S. O. Demokritov} 
\affiliation{Department of Physics, University of Muenster, 48149 Muenster, Germany}

\author{G. de Loubens} 
\affiliation{SPEC, CEA-Saclay, CNRS, Universit\'e Paris-Saclay,
  91191 Gif-sur-Yvette, France}

\author{O. Klein}
\email[Corresponding author:]{ oklein@cea.fr}
\affiliation{Université Grenoble Alpes, CEA, CNRS, Grenoble INP, Spintec, 38054 Grenoble, France}

\date{\today}

\begin{abstract}
This review presents a comprehensive study of the nonlinear transport properties of magnons electrically emitted or absorbed in extended yttrium-iron garnet (YIG) films by the spin transfer effect across a YIG$\vert$Pt interface. Our goal is to experimentally elucidate the pertinent picture behind the asymmetric electrical variation of the magnon transconductance, analogous to an electric diode. The feature is rooted in the variation of the density of low-lying spin-wave modes (so-called low-energy magnons) via an electrical shift of the magnon chemical potential. As the intensity of the spin transfer increases in the forward direction (magnon emission regime), the transport properties of low-energy magnons pass through 3 distinct regimes: \textit{i)} at low currents, where the spin current is a linear function of the electric current, the spin transport is ballistic and determined by the film thickness; \textit{ii)} for amplitudes of the order of the damping compensation threshold, it switches to a highly correlated regime, limited by the magnon-magnon scattering process and characterized by a saturation of the magnon transconductance. Here the main bias controlling the magnon density is thermal fluctuations below the emitter. \textit{iii)} As the temperature under the emitter approaches the Curie temperature, scattering with high-energy magnons starts to dominate, leading to diffusive transport. We find that such a sequence of transport regimes is analogous to the electron hydrodynamic transport in ultrapure media predicted by Radii Gurzhi. This study, restricted to the low-energy part of the magnon manifold, complements part II of this review, which focuses on the full spectrum of propagating magnons.

\end{abstract}
\maketitle

\section{Introduction.} \label{sec:intro}

Diodes are key components in the art of electronics\cite{horowitz1989art}. Their distinctive function is to create an asymmetric electrical conductance that facilitates transport in the forward direction while blocking it in the reverse direction. This asymmetry is exploited in rectification or clipping devices. Controlling the forward threshold voltage is the basis of the bipolar junction transistor. A few of these diodes also offer the unusual feature of negative differential resistance, which is exploited in oscillators and active filters\cite{gunn1963microwave,roy1977tunnelling}. Until recently, it was believed that solid state diodes could only be realized with semiconductor materials. Their band structure allows strong modulation of carrier density by an electrical shift of the chemical potential between the valence and conduction bands.  

Recent advances in the field of spintronics have shown that it is possible to design new types of diode devices that rely on the transport of the electron's spin instead of its charge \cite{Chumak2015,demidov2020spin,barman20212021,chumak2022advances}. In this new paradigm, electrical insulators are good spin conductors by allowing spin to propagate between localized magnetic moments via spin-waves (or their quanta magnons) to carry spin information within the crystal lattice without any Joule dissipation\cite{Cornelissen2015,Goennenwein2015,Thiery2018,kajiwara2010}. The absence of Joule dissipation usually gives the dielectric materials very low magnetic damping, which is associated with a high propensity of the magnons to behave nonlinearly. In ultra-low damping materials, a variation of 0.1\% from the thermal occupancy is usually sufficient to drive the magnetization dynamics into the nonlinear regime\cite{Li2019}. This effect could be exploited for spin diodes \cite{jenkins2016spin,Thiery2018}, spin amplifiers \cite{cornelissen2018spin,Wimmer2019,althammer2021all} or spin rectifiers of microwave signals \cite{tulapurkar05,Sankey2006,tsoi00,chumak2014magnon,harder2016electrical}.

\begin{figure}
    \includegraphics[width=0.49\textwidth]{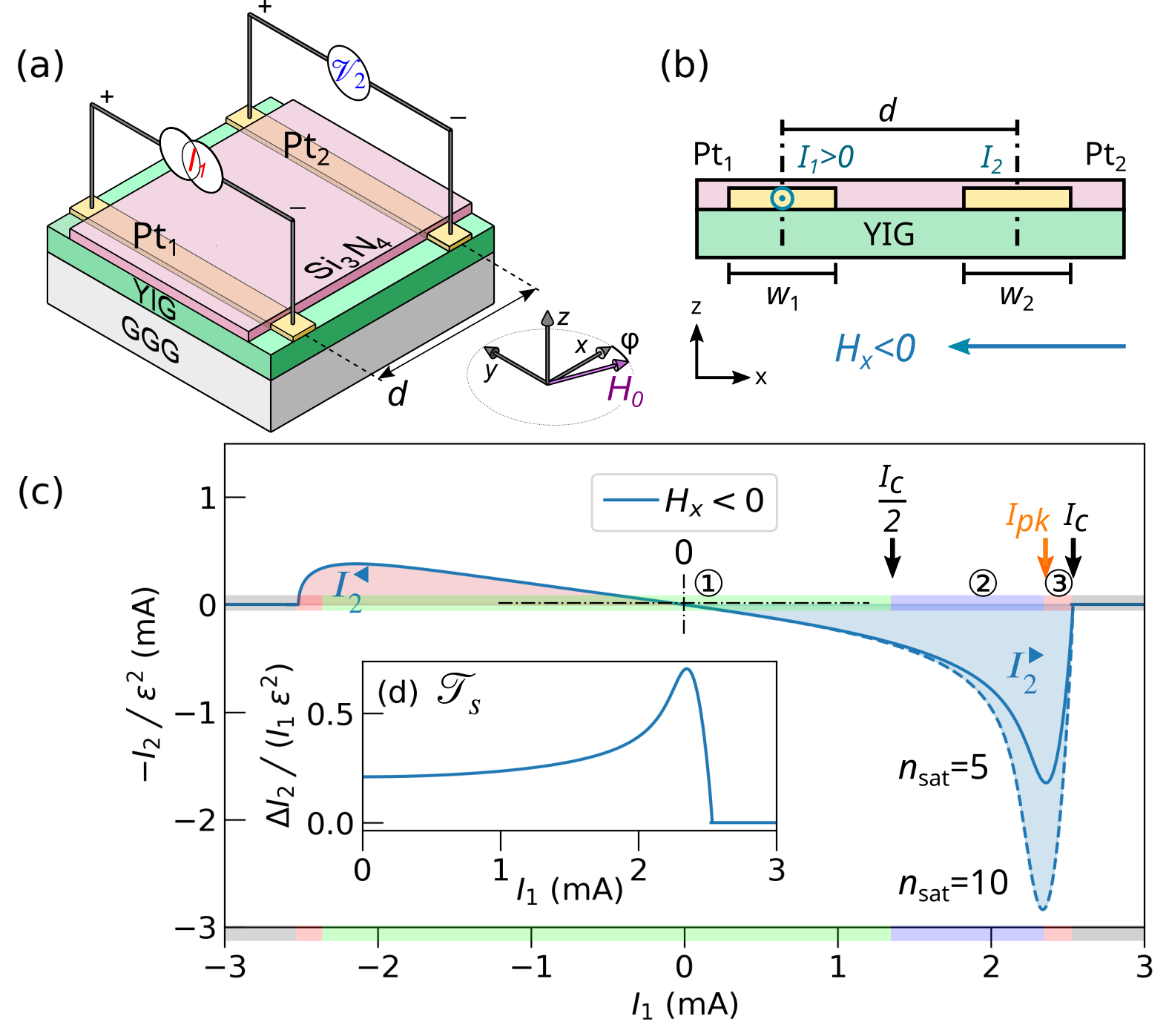}
    \caption{Spin diode effect in a lateral device. (a) Perspective and (b) sectional view of the magnon circuit: an electric current, $I_1$, injected into Pt$_1$ emits or absorbs magnons via the spin transfer effect (STE). The change in density is consequently sensed nonlocally in Pt$_2$ (collector) by the spin pumping current defined as\footnote{The negative sign in front of $I_2$ is a reminder that the origin of the spin Hall effect is an electromotive force whose polarity is opposite to the ohmic losses\cite{Thiery2018a}}, $-I_2 = (\mathscr{V}_2 - \Vbckgnd)/R_2$, where $\Vbckgnd$ is a background signal generated by magnon migration along thermal gradients. (c) Deviation from the nominal thermal occupation of low-energy magnons when $H_x < 0$. We have shaded in blue the deviation in the magnon emission regime corresponding to the forward bias ($\smallblacktriangleright$) and in red the magnon absorption regime corresponding to the reverse bias ($\smallblacktriangleleft$). The inset (d) shows the behavior expressed as a transmission ratio $\mathscr{T}_s \equiv \Delta I_2/I_1$, where $\Delta {I_2} = (I_2^\smallblacktriangleright- I_2^\smallblacktriangleleft)/2$. As shown in (c), an asymmetric nonlinear growth of $I_2$ occurs in the forward bias, corresponding to the so-called spin diode effect.
    However, this growth is limited by $n_\text{sat}$, an effective saturation threshold due to nonlinear coupling between low-energy modes, as expressed by Eq.~(\ref{eq:gamma}). The solid and dashed lines show the behavior for two values of the saturation level, $n_\text{sat}$. The colored current scale in (c) distinguishes 3 transport regimes, \ding{192} (green window): a linear regime, $I_2 \propto I_1$ at low currents, which also includes the reverse bias; \ding{193} (blue window): a nonlinear asymmetric increase above a forward bias of the order of $I_c/2$ \footnote{According to Eq. (\ref{eq:dnk}) and Eq. (\ref{eq:taus}), $\mathscr{T}_K^{-1}$ follows a parabola that intersects the abscissa at $I_\text{th}$ and the current that reaches a 25\% drop \cite{Thiery2018} in signal is the landmark of $I_\text{th}/2$.}; and \ding{194} (red window): above $I_\text{pk}$, a collapse at $I_\text{c}$ of the spin conduction as the temperature of the emitter, $T_1 \rightarrow T_c$. When $\left |I_1 \right | > \left |I_c \right |$ all magnon transport disappears (black window).}
    \label{fig:intro}
\end{figure}

An important milestone in this nascent offshoot of spintronics using electrical insulators, called insulatronics\cite{brataas2020spin}, was the discovery of the interconversion process between the spin and charge degrees of freedom (e.g., by the spin Hall effect\cite{mihai2010current,Miron11,sinova2015spin,Lesne2016,chauleau2016efficient,Vlietstra2013,sanz2020quantification}), which allows the electrical control of $\mu_M$, the magnon chemical potential, via the spin transfer effects (STE) from an adjacent metal electrode\cite{demidov2017chemical,cornelissen2018spin,olsson2020pure,Du2017}. The expected benefit is the realization of a new form of solid state spin diode (see Fig.~\ref{fig:intro}) obtained by electrically shifting $\mu_M$ relative to $E_g$, the energy gap in the magnon band diagram (see Fig.~\ref{fig:be}). The lever-arm is set here by the magnetic damping parameter, $\alpha_\text{LLG}$\cite{hamadeh2012autonomous,hamadeh2014full,Demidov2012,demidov2017magnetization,demidov2017chemical}: the smaller is the damping, the larger is the electrical shift of $\mu_M$ at a given current. Therefore, the ferrimagnet yttrium-iron garnet (YIG), with the lowest magnetic damping known in nature, is expected to lead to the highest benchmark in asymmetric transport performance\cite{Cherepanov1993}. While the spin diode effect in laterally confined geometries is fairly well understood\cite{Slavin09,skirdkov2020spin,finocchio2021perspectives}, its generalization to extended thin films has been largely elusive\cite{Demidov2011}. The corresponding difficulty is the collective dynamics that emerges inside the continuum of magnons when the level splitting between eigenmodes becomes smaller than their linewidth\cite{hamadeh2014full,Collet2016}. Given that magnons are bosons, the predicted appearance of new condensed phases at high powers is of particular interest. These include a superfluid phase produced by a Bose-Einstein condensate (BEC) \cite{Schneider2020,schneider2021control,divinskiy2021evidence,demokritov:06,Tserkovnyak2016,Flebus2016,bender2012electronic}, which could lead to novel coherent transport phenomena \cite{Wimmer2019,bozhko2016supercurrent,kreil2018kinetic,nakata2017spin}.

The manuscript is organized as follows. In the following section, we emphasize the highlights of our transport study observed at high currents on lateral devices deposited on extended thin films. In the third section, we present the relevant analytical framework to account for the electrical variation of the magnon transconductance. This framework is based on a two-fluid model introduced in part II \cite{kohno_2F}, which splits the propagating magnons into either low- or high-energy magnons. To facilitate quick reading of either manuscript,  we point out that a summary of the highlights is provided after each introduction and, in both papers, the figures are organized into a self-explanatory storyboard, summarized by a short sentence at the beginning of each caption. In this first part, we focus mainly on the nonlinear properties that lead to the spin diode effect. It will be shown that these are mainly controlled by the variation of the spin-spin relaxation rates between low-energy magnons. In the fourth section, we present the experimental evidence supporting our interpretation. Finally, we conclude the paper by expressing what we believe to be the interesting future directions of this field. 

\section{Key findings.} \label{sec:key}


The main experimental result presented in this review is that the nonlinear growth of the magnon conductivity observed for the forward bias (magnon emission regime) is rapidly capped by a saturation threshold in the case of extended geometries. Such behavior can be understood by focusing on the nonlinear properties of low-energy magnons, where the limited growth is reminiscent of the saturation effect observed in the ferromagnetic resonance of magnetic films at high power \cite{sparks}. There, the amplitude of the uniform precession mode, the so-called Kittel mode, is constrained not to exceed a maximum cone angle by a power dependent magnon-magnon scattering rate, which increases as the mode amplitude approaches saturation. This process involves the increase in parametric coupling via the magnetostatic interaction between pairs of counter-propagating magnon modes, all oscillating at the Kittel frequency. This process is also known as the Suhl second-order instability or the four-magnon process. We believe that the same process is at work here between our low-energy magnons. We therefore propose a model in which $\Gamma_m$, the magnon-magnon relaxation rate, also increases strongly as $\mu_M$ approaches $E_g$\cite{Demidov2011,demidov2017chemical}. This conclusion is drawn from the experimental observation that in extended thin films the regime $\mu_M \ge E_g$ is never reached despite the use of very large current densities. The fact that it is never reached cannot be attributed to a threshold current, $I_\text{th}$, which would nominally be outside the current range explored. In fact, the current range is large enough to reach the paramagnetic state below the emitter due to Joule heating. Assuming that the damping rate is independent of the current value, the vanishing magnetization is in principle sufficient to bring $I_\text{th}$ into the explored current range, since the threshold current drops to zero at the Curie temperature. Despite this reduction of the magnetization value, we do not detect \emph{nonlocally} any evidence of full damping compensation. We therefore conclude that the increase of $\mu_M$ by the STE is eventually compromised by the concomitant increase of the damping, which continuously pushes $I_\text{th}$ to an unattainable higher level. 
The analytical model proposed to describe the asymptotic approach of $\mu_M$ to $E_g$ is governed by a $\Gamma_m$, whose power dependence is enhanced by a Lorentz factor to limit the growth of mode amplitude below a saturation value. Such an analytical form was first proposed to describe the saturation effects of dipolarly coupled spin waves\cite{suhl:57,suhl:59}. We show that this simple model captures well the high-power transport regime of each of our nonlocal devices, where we have varied the aspect ratio of the electrodes, the magnetic properties, and the film thickness $t_\mathrm{YIG}$ (see Table.~\ref{tab:mat}). The other advantage of this simplified model is that it reproduces the experimental behavior with a very limited set of effective fitting parameters.  

Complementing the transport results with Brillouin light scattering (BLS) experiments, we show experimentally that the asymmetric transport behavior, i.e., the spin diode effect, is indeed predominantly caused by low-energy magnons with energies around $E_K= \hbar \omega_K \approx 30$~$\mu$eV, where $\omega_K/(2 \pi)$ is the Kittel frequency \cite{gurevich2020magnetization}.  This picture is further confirmed experimentally by the enhancement of the spin diode effect in materials with isotropically compensated demagnetization factors, where the uniaxial perpendicular anisotropy $K_u$ compensates the out-of-plane demagnetization field. Here the vanishing small effective magnetization $M_\text{eff} = M_1 - 2 K_u/(\mu_0 M_1) \approx 0$, where $\mu_0$ is the vacuum permeability, reduces the limiting nonlinear magnon-magnon interaction and thus increases the saturation threshold. Nevertheless, even in the case of these thin films, we observe that the growth of the spin diode effect is capped by a saturation threshold, indicating that the rapid growth of $\Gamma_m$ as $\mu_M$ approaches $E_g$ is a generic process. In fact, reducing $M_\text{eff}$ only allows to reduce the static component of the dipolar interaction, but low-energy modes can still interact through the dynamic component of the dipolar interaction.

The picture that emerges from our study of nonlocal magnon transport in extended thin films at high power \footnote{The situation is different under the Pt electrode \cite{divinskiy2021evidence,schneider2021stabilization,merbouche2021,ulrichs2020chaotic}} is thus very different from the \emph{frictionless} many-body condensate around the lowest energy mode, indicating a \emph{reduction} of its relaxation rate. On the contrary, our results indicate an increase in the magnon-magnon relaxation rate, which prohibits the appearance of BEC phenomena. In other words, the reported signal suggests coupled dynamics between numerous modes rather than the BEC picture of a single dominant mode\cite{demokritov:06,Schneider2020,divinskiy2021evidence,schneider2021control}.

Finally, we report a collapse of the magnon transconductance as the emitter temperature, $T_1$, approaches the Curie temperature, $T_c$. In this limit, the population of high-energy magnons becomes significantly larger than the number of polarized spins, spin conduction by low-lying spin excitations is perturbed by collisions with their high-energy counterpart, leading to a sharp decrease in the transmission ratio.

We study how the magnon density enhancement varies with the physical properties of the nonlocal device itself. By systematically varying the transparency of the emitter-collector interface and the thermal gradient near the emitter, we show that the density of low-energy magnons is dominated by thermal fluctuations rather than by the effective damping value, which always remains finitely positive ($> 0$). The process for the forward polarity (magnon emission regime) leads to the finding that the density of low-energy magnons seems to be determined by the temperature of the emitter, $T_1$, with a particle density that increases with increasing temperature. This is in contrast to the usual (ground state) condensation phenomena that normally occur as the temperature decreases. The behavior of our system is similar to that of a free electron gas inside ultrapure 2D materials such as monolayer graphene encapsulated by two layers of hexagonal boron nitride \cite{Bandurin2016,Gurzhi1995}. In the latter, electron transport becomes hydrodynamic at high temperature and can thus be described by the Navier-Stokes equation \cite{Polini2020}. This leads to unusual transport behavior, as first predicted by Radii Gurzhi in the 1960s \cite{Gurzhi1963}. Our results thus suggest that very similar processes are at work in ultra-low damping magnon conductors\cite{Ulloa2019,rodriguez2022probing,sano2023breaking}, the consequence of which is the emergence of a magneto-hydrodynamic regime at high power.

\section{Analytical framework.} \label{sec:analytical}

Since the discovery of spin injection into an adjacent magnetic layer\cite{saitoh:182509}, there have been extensive efforts to describe the high-power regime of magnets excited by STE from an external source. For highly confined geometries, such as nanopillars, the pertinent picture that emerged is that of a dynamical state dominated by a single eigenmode whose damping is reduced or enhanced by the external flow of angular momentum\cite{hamadeh2014full,Demidov2012,demidov2017magnetization,Slavin09}. Its peculiarity is the emergence of an auto-oscillation regime, which characterizes the damping compensation threshold, $I_\text{th}$. The selection of the dominant eigenmode is mostly based on the relaxation rate criterion, which favors the lowest possible value\cite{Slavin09,demidov2017magnetization}. It usually coincides with the Kittel mode, whose characteristic is to have the longest wavelength or smallest wavevector (a criterion mostly valid for strong confinement). In pillars based on ultrathin films, the eigenfrequency is roughly determined by the so-called Kittel frequency\cite{Kittel48}, which has a simple analytical form $\omega_K \equiv \gamma \mu_0 \sqrt{H_0 (H_0 + M_s)}$ for an in-plane magnetized sample, where $\gamma$ is the gyromagnetic ratio. In nanopillars, the degeneracy with other magnons modes is removed by confinement, preventing them from playing a significant role in the dynamics\cite{naletov11,Slavin09}. An additional advantage of working with nanopillars is that the free magnetic layer is efficiently thermalized with the substrate and any Joule effect produced by high current densities is reduced. In the case of current perpendicular to the plane (CPP), the value of the threshold current for damping compensation is set by $I_\text{th} = 2 \Gamma_K e \mathcal{N}/\epsilon_1$, where $\Gamma_K=\alpha_\text{LLG} \omega_K$ is the relaxation rate of the Kittel mode, $\alpha_\text{LLG}$ is the damping parameter, $\mathcal{N} = V_1 M_1 /(\gamma \hbar)$ is the effective number of spins that can flip within $V_1 = t_\text{YIG} \cdot w_1 \cdot L_\text{Pt}$, the magnetic volume fed by the spin emitter, and $\epsilon_1$ is the overall efficiency of both the spin-to-charge interconversion and the interfacial spin transfer process that depends on spin diffusion length of Pt and spin mixing conductance at the Pt$|$YIG interface [see below Eq.~(\ref{eq:epsilon})]\cite{Slavin09,hamadeh2012autonomous}. While the introduction of a well-defined $I_\text{th}$, as described above, successfully allows to describe the high-power regime of STE in confined geometries, where the magnon modes are discrete, it will be shown below that the application to extended thin films, where the magnon dispersion is continuous, requires the additional consideration of coupling between degenerate modes. This is done by introducing a power dependent damping rate leading in this case to a threshold current that increases as $E_g - \mu_M(I_1)$ vanishes.

\subsection{Magnon transport in extended thin films.}

\begin{figure}
    \includegraphics[width=0.49\textwidth]{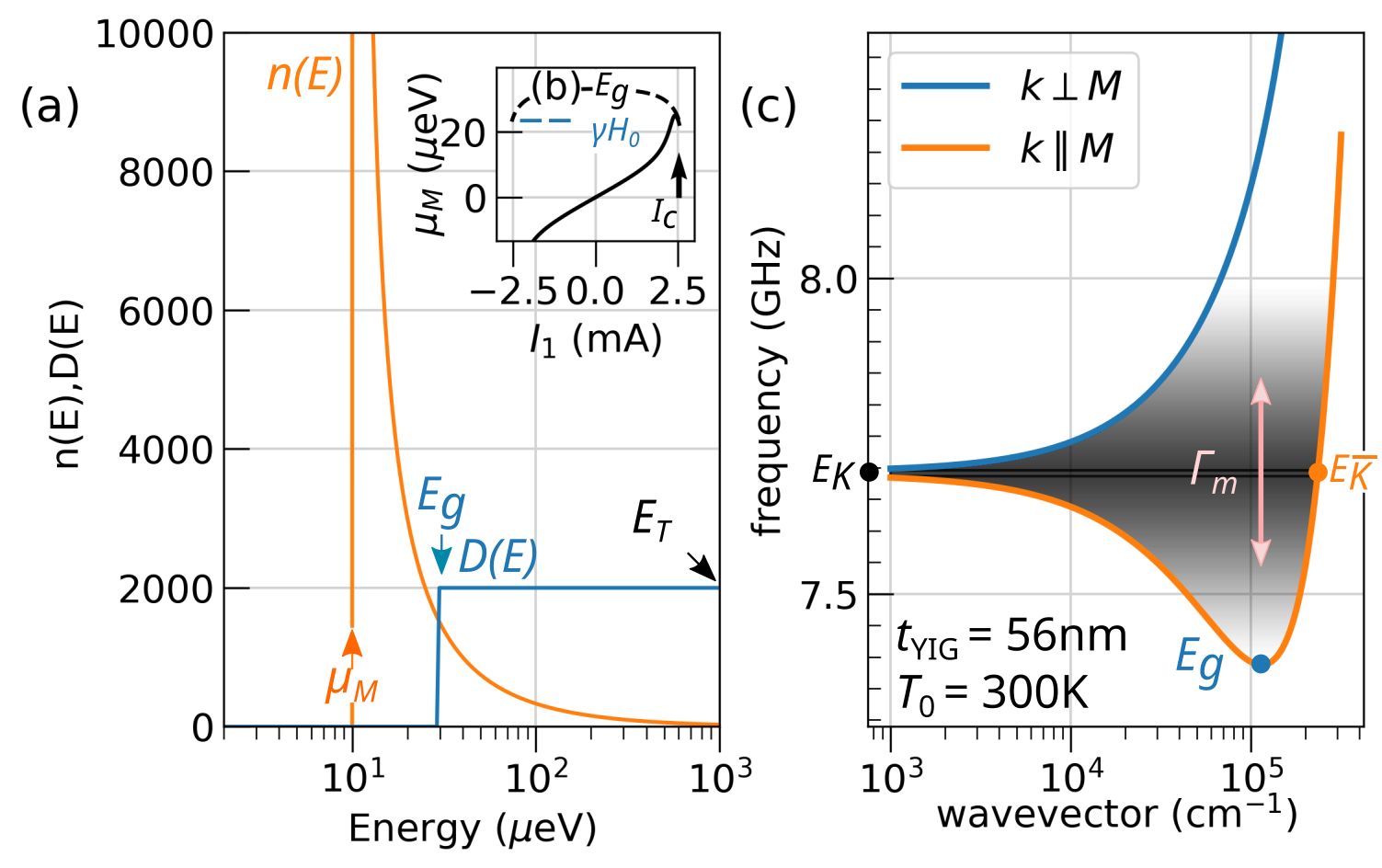}
    \caption{Magnon transport in extended magnetic thin films. (a) At the bottom of the magnon manifold, magnons behave as a two-dimensional gas with a step-like density of states, $D(E)$ (blue). The non-equilibrium magnon distribution, $n(E)$ (orange), can be modulated by shifting $\mu_M$, the magnon chemical potential.  (b) In our case, the shift of $\mu_M$ is produced by an electric current, $I_1$, injected into Pt$_1$ (emitter). (c) Magnon dispersion for different in-plane propagation directions, $\theta_k$. The plot is shown for the YIG$_C$ film at 300~K. The gray shading emphasizes the degeneracy weight of the magnons, i.e., their tendency to excite parametrically degenerate modes. The black dot denotes $E_K$, the energy of the Kittel mode ($k\rightarrow 0$), the blue dot denotes $E_g$, the minimum of the magnon band, and the orange dot denotes $E_{\overline{K}}$, the mode degenerate with $E_K$ with the highest wavevector. The saturation instability described in Fig.~\ref{fig:intro}(b) produces a strong enhancement of $\Gamma_{m}$, the nonlinear coupling between degenerate modes in the energy range $E_K \pm (E_K-E_g)$. This forces $\mu_M$ to asymptotically approach $E_g$ at large currents, as shown in (b).}
    \label{fig:be}
\end{figure}

Extending the zero-dimensional (0D) model \cite{Slavin09,Collet2016,demidov2016direct} to either the one-dimensional (1D) \cite{merbouche2021,an2014control,divinskiy2018excitation} or two-dimensional (2D) \cite{Thiery2018,Evelt2018,Demidov2012} case systems has proven challenging. Experimentally, this extension has been made possible by the discovery of regenerative spin transfer mechanisms, such as the Spin Hall or Rashba effects, which provide STE while allowing the charge current to flow in-plane (CIP). The nonlocal device used to measure magnon transconductance is shown in Fig.~\ref{fig:intro}(a). It has two contact electrodes deposited on YIG and subjected to the spin Hall effect (in our case, two Pt strips $L_\text{Pt}=30$~$\mu$m long, $w_\text{Pt}=0.3$~$\mu$m wide and $t_\text{Pt}=7$~nm thick). The two stripes are separated by a center-to-center distance, $d$. To reach the high power regime, current densities up to $10^{12}$ A/m$^2$ are injected in Pt$_1$. 

To characterize the transport properties, we propose to focus on the dimensionless transmission coefficient $\mathscr{T}_s \equiv I_2/I_1$, which corresponds to the ratio of the emitted and collected spin Hall currents circulating in the two Pt strips, Pt$_1$ and Pt$_2$. The quantity $\mathscr{T}_s/R_1$ can be loosely related to the magnon transconductance \footnote{More precisely, the magnon component requires an additional renormalization by the product $\epsilon_1 \cdot \epsilon_2$, which factors out the interfacial efficiencies of the spin-to-charge interconversion process at both electrodes (see part II\cite{kohno_2F})}.

Note that since STE is a transverse effect, the different cross sections of the spin and charge currents must be taken into account, resulting in the STE efficiency $I_\text{CIP} = I_\text{CPP} \cdot L_\text{Pt}/t_\text{Pt}$. Assuming that all injected spins remain localized below the emitter, one can obtain an estimate for the amplitude of the critical current that compensates for the $\Gamma_K$ attenuation for this confined mode:
\begin{equation}
\dfrac{I_\text{th}} e = 2 \; \dfrac{\Gamma_K} {\epsilon_1} \; \dfrac {t_\text{YIG} \cdot w_1 \cdot t_\text{Pt} \cdot M_1} {\gamma \hbar} \; , \label{eq:ith0}
\end{equation}
where $e$ is the electron charge. A numerical calculation yields $I_\text{th}\approx 1.25$~mA for YIG$_C$ films (see Table.~\ref{tab:mat}). As will be shown below, the corresponding expression for the threshold current in \emph{extended thin films}, $\mathcal{I}_{\text{th}}$, differs significantly from the simplistic estimate above. First, the magnon continuum with the absence of discrete energy levels in unconfined geometries allows for nonlinear interaction between degenerate eignemodes\cite{suhl:57,suhl:57,anderson:55}, whose signature is a power dependent magnon-magnon scattering time\cite{Bauer2015,schultheiss2009direct,schultheiss2012direct,barsukov2019giant}. Second, the Joule heating and the less efficient thermalization of the emitter lead to a significant increase of the emitter temperature $\left . T_1 \right |_{I_1^2} = T_0 + \kappa R I_1^2$ for large currents.  In our notation $\kappa$ represents the thermalization efficiency of our Pt stripe. It is the coefficient that determines the temperature rise per deposited joule power (see Fig.~S1 in \cite{kohno_2F}). Here the rise is produced relative to $T_0$, the temperature of the substrate. This introduces additional complexity due to a significant variation of $M_1$, the magnetization under the emitter. These difficulties are large spectral shifts of the magnon manifold on the scale of $E_K-E_g$: the relevant energy range for large changes in the low-energy magnon density in thin films. In the following we will present a model that includes all these effects and allows to model the magnon transmission ratio [see Fig.~\ref{fig:intro}(c)]. In this model, $\mathcal{I}_{\text{th}}$ depends on a low current nominal estimate of $\mathcal{I}_{\text{th},0}$ whose value is to be extracted from the fit to the data. Although the estimate provided by Eq. (\ref{eq:ith0}) gives the correct order of magnitude, the fit value is systematically larger, indicating that the volume affected by STE is much larger than just the YIG volume covered by the Pt$_1$ electrode.

\subsection{Spin transfer effect in extended thin films.}

\subsubsection{Magnon chemical potential.}

On the analytical side, the single mode picture\cite{Slavin09} is no longer relevant and should be replaced by a statistical distribution filled by an integration over all possible wavevectors $\int d\mathbf{k}/(2\pi)^3$\cite{cornelissen2018spin}. The appropriate framework to describe the out-of-equilibrium regime of the magnon gas is the Bose-Einstein statistics [see Fig.~\ref{fig:be}(a)]: 
\begin{equation}
n(\omega_k) = \dfrac 1 {\exp \left [(\hbar \omega_k - \mu_M)/(k_B T_1) \right ] -1 }, \label{eq:bose}
\end{equation}
where $\mu_M$ is the electrically controlled chemical potential of the magnons whose value follows the expression\cite{demidov2017chemical}:
\begin{equation}
\mu_M = E_g I_1/\mathcal{I}_{\text{th}}, \label{eq:mum}
\end{equation}
where the analytical expression of $\mathcal{I}_{\text{th}}$ in extended thin films will be defined later in Eq.~(\ref{eq:ik}). A drastic change in the density of low-energy magnons is expected when $\mu_M$ approaches $E_g$, indicated by a blue dot in Fig.~\ref{fig:be}(c) \cite{demokritov:06}. This energy level corresponds to the spin-wave eigenmode with the lowest possible energy in the dispersion relation of in-plane magnetized thin films. It occurs for spin-waves propagating along the magnetization direction with wavevectors $k_g \approx 2\pi \cdot 10^5$~cm$^{-1}$, or wavelength $\lambda_g \approx 600$~nm. Such a wavelength is still very large compared to the film thickness, and for all practical purposes we will assume that the magnons behave as a 2D fluid in the lowest spectral part of the magnon manifold as shown in Fig.~\ref{fig:be}(c), and thus the density of states is a step function as shown in Fig.~\ref{fig:be}(a)\cite{jonker2014density}. In addition, in thin films, the gap between $E_g$ and the energy of the Kittel mode, $E_K$ (black dot), is small, so that for all practical purposes $E_g \approx E_K$ can be approximated by the analytical expression of $\hbar \omega_K$. 
It will be shown below that in extended thin films both $\mathcal{I}_{\text{th}}$ and $E_g$ depend on $I_1$, the current bias, for two different reasons: \textit{i)} as mentioned above, the poor thermalization of the electrodes leads to a decrease of $M_1$ with increasing thermal fluctuation produced by Joule heating ($\propto I_1^2$), and \textit{ii)} the finite degeneracy of the magnon bands allows nonlinear coupling between eigenmodes via dipolar coupling [see Fig.~\ref{fig:be}(b)]. As a consequence, $\Gamma_M$, the magnon-magnon relaxation rate which defines $\mathcal{I}_\text{th}$, increases for higher densities of low-energy magnons.

A drastic simplification can be made by noting that for boson statistics there are two ingredients that give rise to a change in the magnon occupation described by Eq.~(\ref{eq:bose}): one is the chemical potential, $\mu_M$; the second is the temperature of the emitter, $T_1$. The former dominantly affects the low-lying spin excitations, while the latter affects the whole spectrum, mostly weighted by the high-energy part. It is therefore natural to simplify the problem as a competing two-fluid problem: one of magnetostatic nature at energy around $E_K$ and a second of exchange nature at energy around $E_T \sim k_B T_1$. The segregation within the two-fluids is precisely the focus of the related work in part II\cite{kohno_2F}. Thus, to describe the magnon transconductance in our nonlocal devices, we propose to split the transmission ratio $\mathscr{T}_s = \mathscr{T}_K + \mathscr{T}_T$ into two separate components, each accounting for the contribution of the low-energy and the high-energy magnon \cite{kohno_2F}.

\subsubsection{Transconductance by low-energy magnons.}

 We now concentrate on deriving an analytical expression for the low-energy magnon transmission ratio $\mathscr{T}_K$ in the linear regime ($I_1 \ll I_\text{th}$). Starting from Eq.~(\ref{eq:bose}), we derive the \emph{linear} variation of the low-energy magnons around the Kittel energy, $E_K$, which is measured by $\Delta n_K = \left | n_K(+I_1) - n_K(-I_1)\right | / 2$, where the number of magnons emitted by the STE ($I_1 \cdot H_x <0$) in Fig. ~\ref{fig:intro}(b) is measured relative to the number of magnons absorbed while reversing the current (or magnetization) direction ($I_1 \cdot H_x >0$)\cite{borisenko2018relation}. In Fig.~\ref{fig:intro}(c) these two biases are symbolized by $\smallblacktriangleright$, for the forward bias (magnon emission), and by $\smallblacktriangleleft$, for the reverse bias (magnon absorption). The subtraction allows to distinguish electrically produced magnons ($n(\omega_k)$ is odd in $I_1$) from magnons produced by pure Joule heating ($n(\omega_k)$ is even in $I_1$)\cite{Cornelissen2015,uchida2008observation,uchida2010insulator,guckelhorn2020quantitative}. Following the expression of $n(\omega_k)$ in Eq.~(\ref{eq:bose}), one obtains an analytical expression for the variation of the number of low-energy magnons\cite{borisenko2018relation}:
\begin{equation}
    \left . \Delta n_K \right |_{I_1}\approx \dfrac {k_B T_1}{\hbar \omega_K} \; \dfrac {I_1} {\mathcal{I}_\text{th}} \dfrac 1 {1 - \left ( I_1 / {\mathcal{I}_\text{th}} \right )^2}\hspace{0.2cm}, \label{eq:dnk}
\end{equation}
where $\mathcal{I}_\text{th}$ was introduced in Eq.~(\ref{eq:mum}). The application of current causes a relative decrease of the magnetization according to the expression $\Delta M_{1} = \Delta n_K \gamma \hbar/ V_1$, where $V_1$ is the effective propagation volume of these magnons. These fluctuations are then detected nonlocally by the change in the spin pumping signal generated in the collector: $\Delta I_2 / e = \epsilon_2 \; \omega_K ( \sigma_\text{Pt} w_2/G_0) \Delta M_2 /M_2 $, where $\epsilon_2$ is the total interfacial efficiency of spin-to-charge interconversion at the collector, $w_2$ is the width of the collector, $G_0 = 2e^2/h$ is the quantum of the conductance, and $M_1$ and $M_2$ are the magnetization values under the emitter and collector, respectively. Since the latter two values differ when the emitter and collector are at different temperatures, we have $\Delta M_2 = \zeta \; M_1 \Delta M_1 /(2 M_2)$, where $\zeta = \text{e}^{-d/\lambda_K} \ll 1$ is the attenuation ratio of the spin signal as magnons propagate from the emitter to the collector, where $\lambda_K$ is the characteristic decay length and the factor 1/2 takes into account that an equal flux of magnons will propagate undetected in the opposite direction. Thus, the transmission ratio of low-energy magnons
\begin{equation}
\left .  \mathscr{T}_K \right |_{I_1} \equiv \dfrac{\Delta I_2}{I_1} \propto  \epsilon_2  \cdot \dfrac{e \omega_K}{I_1} \cdot  \dfrac{M_1}{M_2} \cdot \left . \Delta n_K \right |_{I_1} \label{eq:taus}
\end{equation}
finds an analytical expression, which in the linear regime ($I_1 \ll \mathcal{I}_{\text{th}} $) simplifies as:
\begin{widetext}
\begin{equation}
    \left .  \mathscr{T}_K \right |_{I_1 \rightarrow 0 }  \propto  { \text{e}^{-d/\lambda_K}}  \cdot \epsilon_1 \epsilon_2 \cdot \dfrac{k_B T_1}{\alpha_\text{LLG} \hbar \omega_K} \cdot  \dfrac{\sigma_\text{Pt} w_2} {G_0} \cdot  \dfrac{M_1}{M_2} \cdot \dfrac{\gamma \hbar} {M_1 t_\text{YIG} w_1 t_\text{Pt}}, \label{eq:dim}
\end{equation} 
\end{widetext}
where the value of $\mathcal{I}_{\text{th},0}$ has been replaced by the analytical estimation of $I_\text{th}$ as expressed by Eq. (\ref{eq:ith0}).
The above expression predicts an increase in the magnon transmission ratio with decreasing film thickness $\propto t_\text{YIG}^{-1}$, consistent with recent results\cite{Shan2016}. This follows directly from the fact that the spin pumping signal, $I_2$, is proportional to changes in magnon density. Relating this to an external flow of spins from the interface decaying at a fixed rate, the result simply translates to the concentration being inversely proportional to the magnetic film thickness, all else being equal \footnote{This argument neglects how the changes in the dispersion curve while varying the thickness of the thin film could also affect the transmission ratio.}. This raises the question of how far the transmission can go when $t_\text{YIG}^{-1}\rightarrow 0$. It will be shown below that the answer is disappointing because the nonlocal geometry inherently prevents efficient transport of low-energy magnons. The other geometrical influences are confirmed in the appendix [see Fig.~\ref{fig:widthdep}]. Note that at large currents, Joule heating increases the temperature of the lattice below the emitter, as described in Fig.~\ref{fig:analytic}(a). The corresponding lateral temperature gradients may also introduce additional dependencies\cite{borys2021unidirectional,vogel2018control}, which are not considered in Eq.~(\ref{eq:dim}). Also, we do not consider here the effects of the local gradient of the external magnetic field near the emitter caused by the Oersted field generated by the current $I_1$ \cite{ulrichs2020chaotic}.

\subsection{High power regime.} \label{sec:suhl}

Having established the linear response, this section reviews the phenomena that affect the propagation of low-energy magnons in the strongly out-of-equilibrium regime. We distinguish between thermal effects and nonlinear effects within the low-energy magnon gas. 

\subsubsection{Joule heating.} \label{sec:joule}

\begin{figure}
    \includegraphics[width=0.49\textwidth]{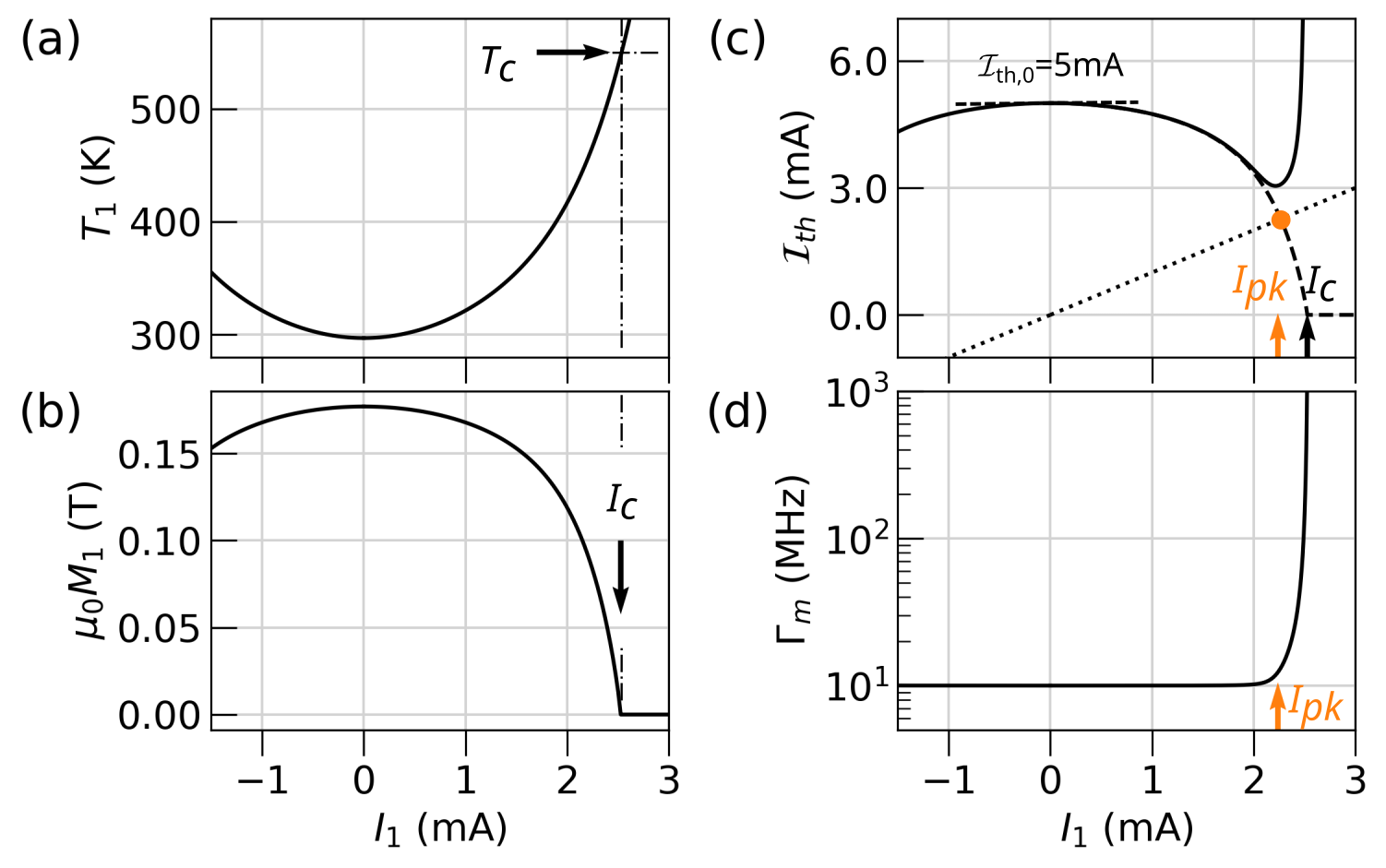}
    \caption{Electrical variation of magnetic properties at high power. (a) Temperature rise at the emitter due to Joule heating, $\left . T_1 \right |_{I_1^2}$. As shown in (b), this results in a reduction of the magnetization below the emitter, $\left . M_{1} \right |_{T_1}$. We define $I_\text{c}$, the critical current to reach $T_c$, the Curie temperature. (c) Variation of the presumed threshold current for damping compensation. The reduction of $M_1$ causes a collapse below $I_\text{c}$ of ${I}_\text{pk} = \mathcal{I}_{\text{th},0} M_{T_\text{pk}} / M_{T_0}$, the expected onset of auto-oscillation for non-interacting magnons. The orange arrow in (c) indicates ${I}_\text{pk}$ when the normal threshold $\mathcal{I}_{\text{th},0} =$ 5~mA. Due to parametric instability, the occupancy of any eigenmode is capped at $n_\text{sat}$ by a sharp increase in $\Gamma_{m}$, the nonlinear coupling between degenerate modes, as shown in (d). This translates in (c) as a sharp increase of $\mathcal{I}_{\text{th}} = \mathcal{I}_{\text{th,0}} \Gamma_{m}/(\alpha_\text{LLG} \omega_K)$.}
    \label{fig:analytic}
\end{figure}

As mentioned above, the 2D geometry prevents efficient thermalization of the magnetic material. Therefore, at high current densities, one should expect a significant variation of $T_1$. We define $T_1=T_0 + \kappa \,R \, I_1^2$, the temperature rise of the emitter caused by joule heating when the substrate is at $T_0$, and again $\kappa$ is the coefficient that determines the temperature rise per deposited joule power. Fig.~\ref{fig:analytic}(a) shows the typical parabolic increase of $T_1$ produced by passing high current densities through a very thin Pt strip of 300~nm width. The collateral damage is a decrease in the saturation magnetization, which decreases with increasing temperature [see Fig.~\ref{fig:analytic}(b)], approximately along the analytical form $M_T \approx M_0 \sqrt{1 - (T/T_c)^{3/2}}$, where $M_0$ is the saturation magnetization at $T=0$~K\cite{kohno_2F,an2021short}. These temperature changes induce a large spectral shift of the magnon manifold. The relevant energy scale here is set by the difference between $E_K-E_g$, which is less than half a GHz in these thin films as shown in Fig.~\ref{fig:be}(c). Furthermore, the rise can easily exceed $T_c$, the Curie temperature, even when trying to use the pulse method as a means to reduce the duty cycle\cite{Thiery2018a}.  We define the current $I_\text{c}$ needed to reach $T_c=T_0 + \kappa R I_\text{c}^2$. In this case, we expect $M_1=0$ at $I_\text{c}$, as shown by the arrow in Fig\ref{fig:analytic}(b). 


In turn, the decrease of $M_1$ should cause a collapse below $I_\text{c}$ of any threshold currents at the emitter site. At this stage we still assume that the magnons remain non-interacting.  We define ${I}_\text{pk}= \mathcal{I}_{\text{th},0} \cdot M_{T_\text{pk}}/M_{T_0}$\cite{borisenko2018relation}, the conjectured threshold current below the emitter heated by the Joule effect, while $\mathcal{I}_{\text{th},0}$ is the nominal threshold current estimated from the magnetic properties at $I_1=0$. In our notation, $T_\text{pk}$ is the temperature of the emitter at ${I}_\text{pk}$, while $T_0$ is its temperature at $I_1=0$. The position of ${I}_\text{pk}$ can be obtained graphically by looking at the intersection of the dashed curve, which represents $\mathcal{I}_{\text{th},0} \cdot M_{T_1}/M_{T_0}$ and a straight line of slope 1 crossing the origin, shown by the dotted points in panel Fig.~\ref{fig:analytic}(c). The vertical orange arrow indicates the expected position of the auto-oscillation onset, assuming that the nominal value is $\mathcal{I}_{\text{th},0}=5$~mA \footnote{This corresponds to about a 4-fold improvement over the estimate by Eq. (\ref{eq:ith0}) for YIG$_C$}. It is important to note that the position of ${I}_\text{pk}$ is very weakly dependent on the nominal value if $\mathcal{I}_{\text{th},0} \gg I_c$ because of the rapid decrease of $M_1$ near $I_\text{c}$ [see Fig.~\ref{fig:analytic}(b)]. We will return to this observation when discussing below the relevant bias to renormalize the data. 

The important conclusion at this stage is that, taking into account the Joule heating, the damping compensation should always be achieved within the range $[-I_c,I_c]$, regardless of the value of $\mathcal{I}_{\text{th},0}$. Thus, as long as the single mode picture remains valid, one should always observe a divergent increase of the number of magnons within the currently explored range. We will show below that this is not the case, and that the culprit is the increased magnon-magnon scattering, which prevents the divergent growth of the magnon density.

\subsubsection{Self-localization effect.}

We begin this discussion by emphasizing a trivial observation about the magnon dispersion curve, which concerns nonlocal devices, i.e., magnon transport outside the volume below the Pt$\vert$YIG interface. Lateral geometries are \emph{not} appropriate to study condensation, which selectively favors the lowest lying energy mode, such as BEC. For in-plane magnetized thin films, the minimum in the dispersion relation, $E_g$, corresponds to a mode with vanishing group velocity, i.e., a non-propagating mode\cite{Kalinikos1986}: cf. blue dot in Fig.~\ref{fig:be}(c). This means that nonlocal devices are inherently insensitive to changes in the magnon population that occur in a localized mode. This situation is exacerbated when changes in bias or design end up increasing the magnon concentration because of the nonlinear redshift of the magnon spectrum. This pushes the entire spectral range of low-lying spin fluctuations at the emitter below the magnon bandgap for outside the emitter. This prevents these magnons from reaching the collector and further promotes self-localization\cite{ulrichs2020chaotic,Slavin2005b,schneider2021stabilization}. The latter is further enhanced by temperature variations caused by Joule heating when large currents are circulated in the emitter. 

These effects can be mitigated by lowering the nonlinear frequency shift \cite{Soumah2018,divinskiy2019controlled,Evelt2018,guckelhorn2021magnon}. A first possibility is by tilting the sample out-of-plane. There is a peculiar angle where the depolarisation effect vanishes\cite{Vonsovskii1966}.  A second possibility is to use a material whose uniaxial anisotropy, $\mu_0 K_u$, compensates for the out-of-plane demagnetization factor, $\mu_0 M_s$, leading to a vanishing effective magnetization. Note that full redshift cancellation requires tuning the uniaxial anisotropy to a precision of the order of $(E_K-E_g)/(\gamma \hbar)$ (see below). In this case, the Kittel frequency simply reduces to $\omega_K = \gamma H_0$ and is independent of the magnetization amplitude or direction. It has already been noted that this eliminates the self-localization effect of the depolarization factor and thus promotes spin propagation outside the emitter region\cite{Evelt2018}. It will be shown in part II that even if the redshift is extinguished and the product $\epsilon_1 \cdot \epsilon_2 \ll 1$ is omitted, the transmission ratio remains well below 50\%. This upper limit is actually expected, considering that less than half of the magnons propagate in a direction captured by the collector. Finally, we add that this difficulty of efficient transmission affects not only low-energy magnons, but also high-energy magnons, which suffer from very short decay lengths (see part II).

\subsubsection{Lorentz factor enhanced magnon-magnon decay rate.}

We now focus on the inter-magnon nonlinearity that occurs at high power. We are mainly concerned with saturation effects. This instability arises from the non-isochronous precession of the magnetization (inherent to elliptical orbits), which radiates at harmonics of the eigenfrequency, allowing parametric excitation of other modes \cite{Duan2014}. This problem was first described by Harry Suhl \cite{Suhl1957,suhl:59}. It is interpreted that the number of magnons that can fill a particular mode is limited by its nonlinear coupling with other magnon modes, by introducing a magnon-magnon relaxation time that depends on the mode occupation.
The usual requirement is to find a degenerate eigenmode within the linewidth. Since this effect depends on the level of degeneracy, it becomes dominant in extended thin films due to the increase in mode density. Moreover, the peculiar shape of the band structure of the magnons at low energy levels introduces a discrimination between the different frequencies in terms of the number of degenerate modes. It turns out that the energy level with the largest number of degenerate modes occurs precisely at the Kittel frequency. This is emphasized in Fig.~\ref{fig:be}(c) by shading the degeneracy weight in gray. 
It should also be noted that among all the modes being degenerate at $E_K$, the mode with the highest group velocity, which also propagates along the normal direction to the wires, is the point $E_{\overline{K}}$ marked by an orange dot in Fig.~\ref{fig:be}(c). Interestingly, the wavelength at the orange dot here is of the order of $1/w_1$, the lateral size of the Pt$_1$ electrode. This suggests that low-energy magnon transconductance is preferentially carried by magnons at this particular position in the dispersion curve. 

To describe the nonlinear interaction between magnons we introduce a saturation occupancy $\mathcal{N}_\text{sat}$, which marks the maximum number of magnons that one can put in one mode before decay to degenerate energy levels starts to kick in. Near this threshold, we assume that the damping rate follows the equation\cite{Loubens2005}:
\begin{equation}
\left . \Gamma_m \right |_{I_1}= \dfrac {\Gamma_K} {\sqrt{1-\left ( \left . \Delta n_K \right |_{I_1}/ \mathcal{N}_\text{sat} \right )^2}} , \label{eq:gamma}
\end{equation}
with $\Gamma_m - \Gamma_K$ representing the nonlinear enhancement of the relaxation rate caused by magnon-magnon scattering.  The dependence of  $\Gamma_m$ on $I_1$ is plotted in Fig.~\ref{fig:analytic}(d).

\subsubsection{Current threshold in extended thin films.}

To account for the increase of correlation between magnons discussed above, we replace $I_\text{th}$ in Eq.~(\ref{eq:ith0}) by
\begin{equation}
  {\left .  \mathcal{I}_\text{th}  \right |_{I_1}}= \mathcal{I}_{\text{th},0} \dfrac{M_{T_1}}{M_{T_0}}  \dfrac{\left . \Gamma_m \right |_{I_1}}{ \Gamma_K}   . \label{eq:ik}
\end{equation} 

Introducing this new expression of $\mathcal{I}_\text{th}$ into Eq.~(\ref{eq:dnk}) gives a transcendental equation for $\mathcal{I}_\text{th}$, whose dependence on $I_1$ is shown in Fig.~\ref{fig:analytic}(c). As $\Delta n_K(I_1)$ approaches $\mathcal{N}_\text{sat}$ by increasing $I_1$, $\mathcal{I}_\text{th}$ rises sharply together with the damping $\Gamma_m(I_1)$ as shown in Fig.~\ref{fig:analytic}(d) according to Eq.~(\ref{eq:gamma}-\ref{eq:ik}).  The consequence of this increase is that $\mathcal{I}_\text{th}$ remains unreachable due to a redistribution of the injected spin among an increasing number of degenerate eigenmodes. The consequence for the dependence of $\mu_M$, the chemical potential of the magnons, on $I_1$ is shown in Fig.~\ref{fig:be}(b). Near the origin, the linear dependence of $\mu_M$ on $I_1$ is set by the intrinsic damping parameter. As Joule heating begins to decrease the magnetization $M_1$ below the emitter, it shifts the curve upward. In the same way, the decrease in magnetization pushes $E_g$ to a lower energy in accordance with the red-shift nonlinear frequency coefficient. The sharp rise in $\Gamma_m(I_1)$ near $I_\text{c}$ stops the rise and $\mu_M$, which eventually approaches $E_g$ asymptotically at high currents. 

Fig.~\ref{fig:intro}(c) shows the expected behavior produced at $I_2$ for two values of $n_\text{sat} =5$ (solid line) or 10 (dashed line). Here we use as parameter $n_\text{sat} = \mathcal{N}_\text{sat} / \mathcal{N}_\text{NL} $ the value expressed relative to $ \mathcal{N}_\text{NL} = \mathcal{N} \Gamma_K / \omega_M$, which marks the onset when the change in magnetization becomes of the order of the linewidth, with $\omega_M = \gamma \mu_0 M_1 \approx 2\pi \times 4.48$~GHz \cite{gurevich2020magnetization}. In these data we assume that $I_c=2.5$~mA and $I_\text{pk}=2.2$~mA, which is equivalent to assuming that $\mathcal{I}_{\text{th},0}=5$~mA [see Fig.~\ref{fig:analytic}(c)]. 

Depending on the current values, we observe 3 regimes of transport: \ding{192}: $I_1 < I_c/2$, where we have a linear behavior $I_2 \propto I_1$; \ding{193}: $I_1 \in [{I}_c/2,{I}_\text{pk}]$, where we have an asymmetric polynomial increase $I_2 \propto ( 1-I_1^2/I_c^2 )^{-1}$, the regime for the spin diode effect in an extended film; \ding{194}: $I_1 \in [I_\text{pk},I_c]$, where we have a drop of $I_2 \propto (1 - I_1^3/I_c^3)^{1/2}$. In the following we will use Eq.~(\ref{eq:dnk}) combined with Eq.~(\ref{eq:ik}) to fit the data. Note that the result is obviously inverted by reversing the field direction (not shown). 
 
We conclude this section by emphasizing that the susceptibility of low-energy magnons to capture external angular momentum flux in a thermally changing environment predicts a distinctive nonlinear shape for the current dependence of the magnon transmission ratio, which will be confirmed by experimental data in the following sections.

\section{Experiments.} \label{sec:experiment}

In this section, we present the experimental evidence supporting the physical picture presented above. We focus on the nonlinear and asymmetric transport properties, our so-called spin diode effect, and the extraction of the relevant parameters that govern it.

\begin{figure}
    \includegraphics[width=0.49\textwidth]{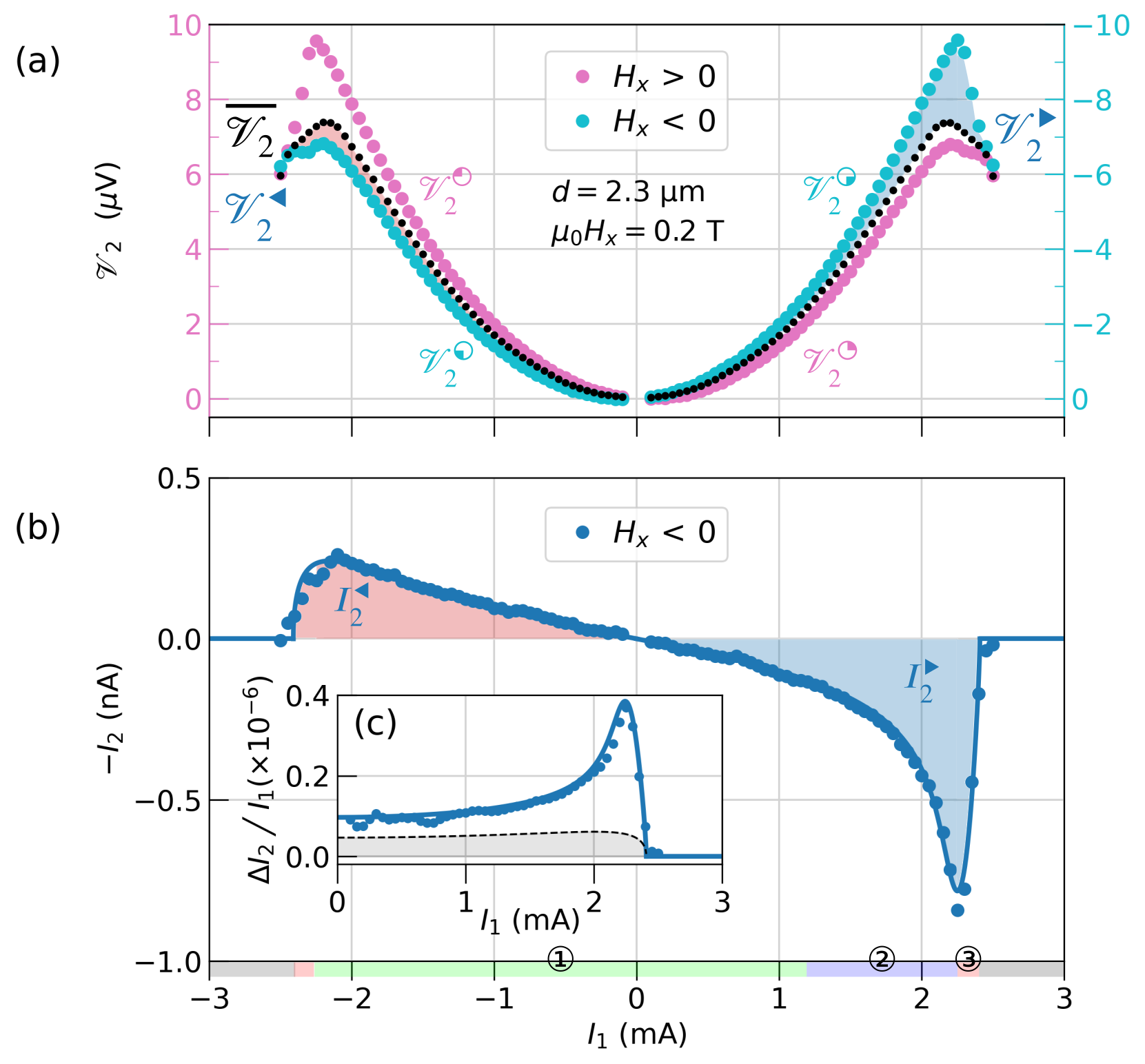}
    \caption{Experimental observation of the spin diode signal. Panel (a) shows the $I_1$ dependence of the nonlocal voltage $\mathscr{V}_2 = (V_{2,\perp} - V_{2,\parallel})$: voltage difference between the normal and parallel configuration of the magnetization with respect to the direction of current flow. The data are shown for both positive (left axis in pink) and negative (right axis in cyan) polarity of the applied magnetic field, $H_x$. The sign of $\mathscr{V}_2$ is reversed when the field is inverted. The nonlocal voltage $\mathscr{V}_2 = -R_2 I_2 + \Vbckgnd$ is decomposed into an electric signal, $I_2$, and a thermal signal, $\Vbckgnd$ (black). Panel (b) shows the variation of $I_2$ at $H_x <0$ in forward and reverse bias. The blue and red shaded areas in panels (a) and (b) highlight respectively the regimes of magnon emission and absorption respectively for the forward and reverse bias. Panel (c) plots the magnon transmission ratio $\mathscr{T}_s= \Delta I_2/I_1$, where $\Delta {I_2} \equiv (I_2^\smallblacktriangleright- I_2^\smallblacktriangleleft)/2$.  The solid line is a fit with Eq.~(\ref{eq:dnk}), (\ref{eq:gamma}) and (\ref{eq:ik}) using $n_\text{sat}=4$ and  $\mathcal{I}_{\text{th},0}=6$~mA. The black shaded region shows the assumed background contribution to spin conduction by high-energy magnons, $\Sigma_T$ (cf.\cite{kohno_2F}).  The data are collected on the YIG$_C$ thin film at ambient temperature, $T_0=300$~K,  using a device operating in the long-range regime ($d=2.3$~$\mu$m).}
    \label{fig:raw}
\end{figure}

 \begin{table*}[t]
  \caption{Physical properties of the magnetic garnet films (values at $T_0=300$~K).}
  \begin{ruledtabular}
    \begin{tabular}{*{11}{c}}
     & {$t_\text{YIG}$  (nm)}  & 
      {$\mu_0 M_s$  (T)}  &  {$H_{Ku}$ (T)} & $T_c$ (K) &
      {$\alpha_\text{YIG} \,\,(\times 10^{-4})$} &
      {$\rho_\text{Pt}$ ($\mu \Omega$.cm)} & {$t_\text{Pt}$  (nm)}  & {$\kappa_\text{Pt}$ (K)} &
      {$G_{\uparrow \downarrow}$ ($\times 10^{18}$ m$^{-2}$)} & $\epsilon$ ($\times 10^{-3}$) \\ \hline
      YIG$_A$ &
      {19} & 
      {0.167} & {+0.005} & {545} & 
      {$3.2$}  & 
      {27.3} & 7 & 480 &
      {0.6} & {2.1} \\ 
      (Bi-)YIG$_B$ &
      {25} & 
      {0.147} & {+0.174} & {560}  &
      {$4.2$}  & 
      {42.0} & 6 & $890$ &
      {$2.4 $} & 7.8\\
      YIG$_C$ &
      {56} & 
      {0.178} & {-0.001} & {544} &
      {$2.0$}  & 
      {19.5} & 7 & 476 &
      {$\circ$:}  
      {$0.64  $} / {\textbullet:}  {$1.9 $} & {$\circ$:} {$1.3  $} / {\textbullet:}  {$3.6$} \\
\end{tabular}
\end{ruledtabular}\label{tab:mat}
\end{table*}

\subsection{Nonlocal magnon transport.} \label{sec:magneto}

\subsubsection{Measurement of the magnon transconductance.}

The experiment is performed here at room temperature, $T_0=300$~K, on a 56~nm thick (YIG$_C$) garnet thin film whose physical properties are summarized in Table.~\ref{tab:mat}. As explained in part II, we deliberately choose a device with a large separation between the two Pt electrodes ($d=2.3$~$\mu$m) to allow the low-energy magnons to dominate the transport properties. By injecting an electric current $I_1$ into Pt$_1$, we measure a voltage $V_2$ across Pt$_2$, whose resistance is $R_2$. To subtract all non-magnetic contributions, we define the magnon signal $\mathscr{V}_2 = (V_{2,\perp} - V_{2,\parallel})$ as the voltage difference between the normal and parallel configuration of the magnetization with respect to the direction of current flow in Pt.  Fig.~\ref{fig:raw}(a) shows the measured variation of $\mathscr{V}_2$ for a large span of $I_1$. The maximum current injected into the device is about 2.5~mA, corresponding to a current density of $1.2\cdot 10^{12}~$A/m$^2$. The Joule heating at this intensity is large enough to reach $T_c$, the Curie temperature. The resulting voltage $\mathscr{V}_2$ is shown in Fig.~\ref{fig:raw} for both positive ($H_x$ pointing to $+x$) and negative ($H_x$ pointing to $-x$) polarity of the applied field, whose amplitude is $\mu_0 H_0=0.2$~T. 
In Fig.~\ref{fig:raw}(a), the expected inversion symmetry with respect to the field polarity has been folded between $H_x > 0$ (left ordinate label in pink) and $H_x <0$ (right ordinate label in blue) to directly emphasize the STE induced deviation between magnon emission and absorption. The measured signal decomposes into two contributions $\mathscr{V}_2 = -R_2 \left . I_2 \right |_{I_1} + \left . \Vbckgnd \right |_{I_1^2}$: one is $I_2$, the electrical signal produced by the STE, and the other is $\Vbckgnd$, a background voltage associated with magnon transport along thermal gradients. The latter voltage corresponds to the Spin Seebeck Effect (SSE). One expects $\Vbckgnd \approx 0$ in well thermalized devices. We emphasize the minus sign in front of $I_2$, which accounts for the fact that the spin-charge conversion is an electromotive force, so the current flows in the opposite direction to the voltage drop. It is thus a reminder that the resulting polarity is opposite to the ohmic losses\cite{Thiery2018a}. In the linear regime ($I_1 \rightarrow 0$), the electrical signal is even/odd with the polarity of $H_x$ or $I_1$, while the thermal signal is always odd/even with $H_x$ or $I_1$\cite{Thiery2018}.

\begin{figure}
    \includegraphics[width=0.49\textwidth]{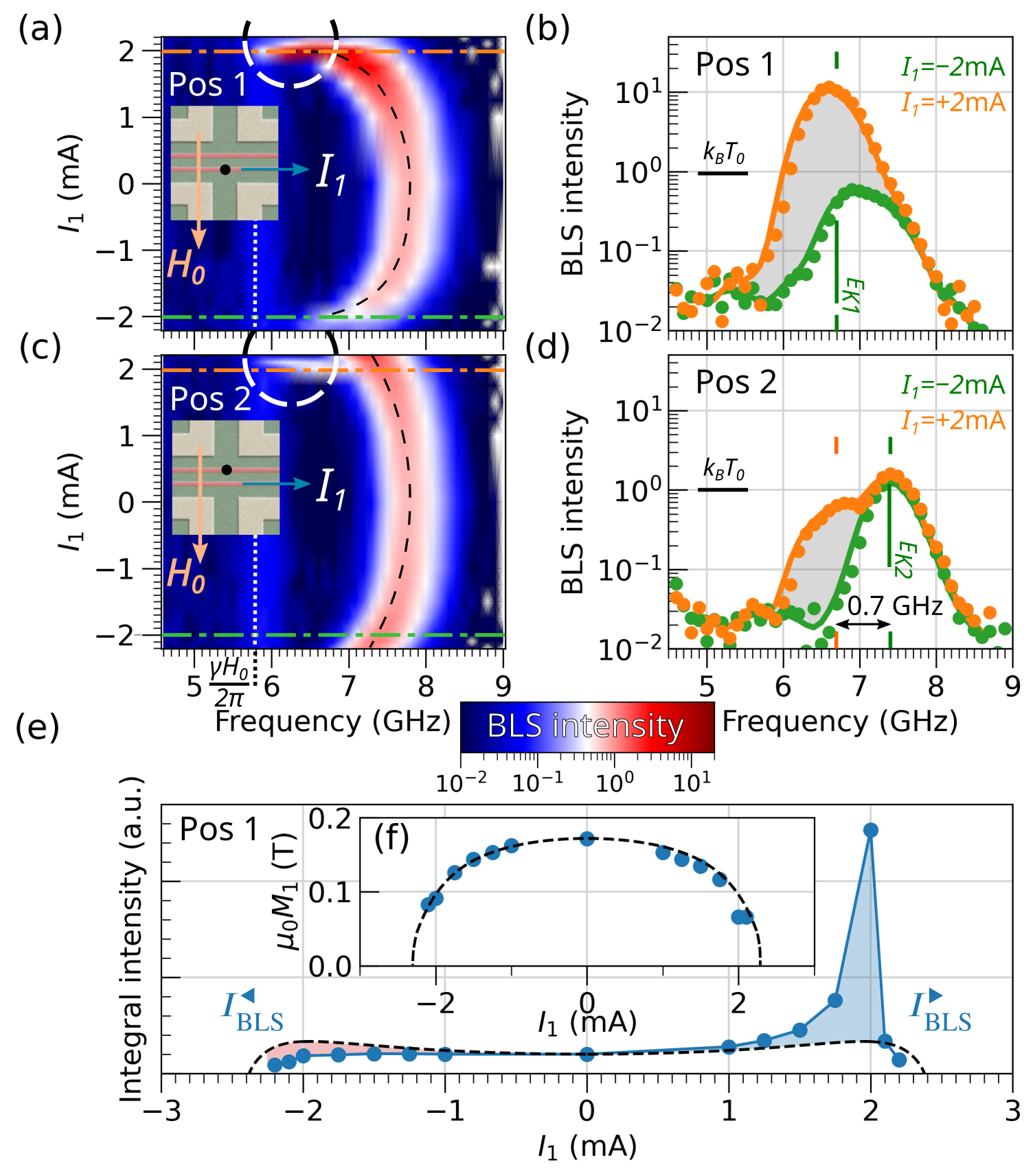}
    \caption{Experimental evidence that the spin diode effect is dominated by low-energy magnons. Panels (a) and (c) show the microfocus Brillouin light scattering ($\mu$-BLS) spectra as a function of $I_1$ performed under the two Pt electrodes (see black dot on the inset) at Pos1 (emitter) and Pos2 (collector) for $H_x <0$. Each BLS spectrum is renormalized by the amplitude of the Kittel peak at $I_1=0$. Panels (b) and (d) show two sections at fixed current $I_1=\pm 2$~mA near $I_\text{pk}$. The green marker indicates the spectral position of $E_K$ at $I_1=2$~mA. The spectral position of the self-localized spin fluctuations, $E_{K1}$, (see circles) at $I_1=+2$~mA (see orange marker) occur well below $E_{g2} \approx E_{K2}$ at Pos2 (see green marker). The shift $E_{K1}-E_{g2}$ is about 0.7~GHz. The paramagnetic limit, $\gamma H_0/(2 \pi)$, is indicated by the yellow vertical dotted line at 5.8~GHz. Panel (e) shows the integrated BLS intensity at Pos1 as a function of $I_1$. The solid blue line is a guide for the eye. The black dashed line shows the expected variation of thermally activated low-energy magnons produced by Joule heating. In echo with Fig.~\ref{fig:raw}, the blue and red shaded areas highlight the variation of magnon emission and absorption with respect to thermal fluctuation. The inset panel (f) shows the evolution of $M_{1}$ induced by Joule heating. The blue points are inferred from the evolution of the Kittel frequency (dashed parabola in panel (a)). The dashed lines are inferred from the SQUID measurements shown in Fig.~S1 of Ref.~\cite{kohno_2F}.  The data are collected on the YIG$_A$ thin film at $T_0=300$~K using a device with a distance $d=1.0$~$\mu$m between the two Pt electrodes.}
    \label{fig:bls}
\end{figure}

Focusing on the nonlinear behavior observed for the $H_x < 0$ configuration (blue data), \footnote{results for the opposite polarity are derived by mirror symmetry}, we label the forward bias as $\mathscr{V}_2^\smallblacktriangleright$ as the nonlocal voltage for $I_1 > 0$ and the reverse bias as $\mathscr{V}_2^\smallblacktriangleleft$ as the nonlocal voltage for $I_1 < 0$. The cancellation of Joule effects can be obtained simply by calculating the difference $\mathscr{V}_2^\smallblacktriangleleft - \mathscr{V}_2^\smallblacktriangleright$, which eliminates any contribution that is even in current. The latter quantity represents the number of magnons produced by the STE relative to the number of magnons absorbed by the STE at a given current bias $|I_1|$. The inset (c) of Fig.~\ref{fig:raw} shows the observed averaged behavior, $\Delta {I_2} \equiv (I_2^\smallblacktriangleright- I_2^\smallblacktriangleleft)/2 = (\mathscr{V}_2^\smallblacktriangleleft- \mathscr{V}_2^\smallblacktriangleright)/(2 R_2)$, expressed as a renormalized quantity $\mathscr{T}_s = {\Delta I_2}/I_1$ , a strictly positive parameter, as suggested by Eq. ~(\ref{eq:dim}). However, the subtraction operation does not allow to separate the behavior between the forward and the backward direction. In fact, the asymmetry of the electrical signal cannot be obtained from the transport data alone. It requires additional input information. This will be provided below by the measurement of the integral intensity of the BLS signal, which directly monitors the low-energy part of the magnon spectrum. Based on the experimental BLS observation in Fig.~\ref{fig:bls}(e), we expect the magnon transmission ratio to show continuity over the origin ($I_1=0$) and to behave as a step function of amplitude in the reverse bias up to $I_\text{c}$. This leads to the following expression $\mathscr{T}_s^\smallblacktriangleleft \approx \left . \mathscr{T}_s \right |_{I_1 \rightarrow 0} T_1/T_0$ for the magnon transmittance in the reversed bias below $I_c$. The ratio $T_1/T_0$ takes into account that the number of thermally excited low-energy magnons in the reverse polarization varies with $I_1$ due to Joule heating. This introduces a second order distortion which will be discussed in more detail in the next section and in part II. So we construct $\Vbckgnd = -\mathscr{V}_2^\smallblacktriangleleft + R_2 \mathscr{T}_s^\smallblacktriangleleft \cdot I_1 $ in the reverse bias from the opposite magnetic configuration. We then force $\Vbckgnd$ to be even in current to get the forward bias behavior. The result is shown as black dots in Fig.~\ref{fig:raw}(a). We can then derive $I_2^\smallblacktriangleright = ( \Vbckgnd - \mathscr{V}_2^\smallblacktriangleright )/R_2$ and $I_2^\smallblacktriangleleft = ( \Vbckgnd - \mathscr{V}_2^\smallblacktriangleleft )/R_2$, which is shown in Fig. ~\ref{fig:raw}(b).

In Fig.~\ref{fig:raw}(b) we observe 3 different transport regimes as predicted above. We have \ding{192} with $I_1 \in [-2,1]$~mA: the spin conductance is approximately constant; \ding{193} with $I_1 \in [1,2.2]$~mA: the spin conductance increases gradually and saturates quickly; and \ding{194} with $I_1 \in [2.2,2.5]$~mA: the spin conductance decreases abruptly to vanish at $I_\text{c}$. Only the regime \ding{192} is anti-symmetric in current. We emphasize at this stage that the sequence of behavior is reminiscent of the 3 regimes predicted in Fig.~\ref{fig:intro}(c). We can also repeat the same analysis to construct the data when $H_x > 0$ (see part II). As expected, the polarity of the asymmetry reverses when the magnetization direction is changed.  In all cases, the forward regime occurs only when $I_1 \cdot H_x <0$, which corresponds to the polarity of the damping compensation. 

In the next section, we will confirm the above behavior using BLS spectroscopy. Throughout the rest of the paper, we will always discuss the data as shown in Fig.~\ref{fig:raw}(c) in the form of $\Delta I_2$ as a function of $|I_1|$ and properly renormalized by $I_1$ and $T_1$ to allow comparison between different devices (see part II \cite{kohno_2F}). 

\subsubsection{Spin diode effect.}

We now discuss in more detail the amplitude of the asymmetric rise of the $I_2$ signal in Fig.~\ref{fig:raw}(c). A striking feature of Fig.~\ref{fig:raw}(c) is the limited growth of the spin diode signal, which is capped by a meager factor of 3 rise compared to the value at small currents. This variation is significantly smaller than the changes in cone angles observed at the damping compensation threshold in nanopillars, which reach several orders of magnitude. Such inefficiency is further confirmed by previous reports aiming at modulating the transport by damping compensation with an additional heavy metal electrode placed between the emitter and the collector, which increases the conduction by a factor of 6 at most \cite{cornelissen2018spin,liu2021electrically,Wimmer2019,guckelhorn2021magnon,althammer2021all}. As shown in Fig.~\ref{fig:be}, this cannot be explained by a threshold current larger than the current explored window $[-I_c,+I_c]$, but rather indicates a strong coupling between the magnons that prevents a large growth of the magnon density. The asymmetric contribution is well accounted for by Eq.~(\ref{eq:dnk}). The best fit is obtained by using $T_c^\star=515$~K, $\mathcal{I}_{\text{th},0}=6$~mA and $n_\text{sat}=4$ and is indicated in the plot by the solid line. The low-energy magnon contribution is added to a background indicated by the dashed line. This background accounts for the competing contribution of high-energy thermal magnons to the electrical transport. Its origin and analytical expression can be found in Ref.~\cite{kohno_2F}. In our fit, it represents a 50\% additional contribution $\Sigma_T/(\Sigma_K+\Sigma_T)=0.5$ in Eq.~(4) of ref.\cite{kohno_2F}.  We also note that the value of $T_c^\star$ used for the fit is significantly different from the Curie temperature of this film (see Fig.~S1 of ref.\cite{kohno_2F}). This point will be investigated in more detail in part II\cite{kohno_2F}.

\subsection{Brillouin Light Scattering.} \label{sec:bls}

It is useful at this stage to confirm spectroscopically that the spin diode effect reported above is indeed due to an asymmetric modulation of the low-energy magnons. BLS is a technique of choice for this purpose, since it is specifically designed to monitor spectral shifts in the magnon population at GHz energies. Furthermore, its high sensitivity allows the detection of fluctuations down to the thermal level. While in the past we have performed comparative studies of the transport and BLS behavior on exactly the same device \cite{Thiery2018,Thiery2018a}, in this particular case we will introduce in the discussion a thinner YIG$_A$ garnet film, whose physical properties are also given in Table \ref{tab:mat}. The change of samples is not intentional and is purely related to the chronological context of the experiments. For all practical purposes, the only relevant difference is a change in the value of $I_\text{c}$ from $I_c=2.5$~mA to 2.1~mA for YIG$_C$ and YIG$_A$, respectively, due to differences in Pt resistivity. We have verified that all other properties discussed here are generic to both samples. 

Since we are interested in local changes in the magnon population, we use micro-focus Brillouin light scattering ($\mu$-BLS) spectroscopy \cite{demidov2015magnonic}. The probe light with a wavelength of 532 nm and a power of 0.1~mW is generated by a single-frequency laser with a spectral linewidth of $< 10$~MHz. It is focused through the sample substrate into a submicron diffraction-limited spot using a 100$\times$ corrected microscope objective with a numerical aperture of 0.85. The scattered light was collected by the same lens and analyzed by a six-pass Fabry-Perot interferometer. The lateral position of the probe spot was controlled using a custom-designed high-resolution optical microscope. The measured signal (the BLS intensity at a given magnon frequency) is proportional to the spectral density of the magnons at that frequency and at the position of the spot. By moving the focal spot across the film surface, we can obtain information about the spatial variations of the magnon spectral distribution. Note that only low-energy magnons contribute to the BLS intensity, since the BLS technique is sensitive only to the wavevectors smaller than $2.4 \cdot 10^{5}$ cm$^{-1}$. However, since the frequency of magnetostatic magnons also depends on the total number of magnons in the sample, the spectral position of the Kittel peak allows to derive information about the local temperature from the BLS data\cite{An2016,birt2013brillouin}.   

Fig.~\ref{fig:bls} compares the current modulation of the spectral occupation below the emitter and collector electrodes labeled Pos1 ($x=0$, emitter) and Pos2 ($x=+d$, collector). The position of the spot is indicated by a black circle in the inset images in Fig.~\ref{fig:bls}(a) and (c). To allow a quantitative comparison of the different locations, all curves have been renormalized by the value of the thermal fluctuations at $I_1=0$, which is assumed to be constant throughout the thin films. In all these measurements, the field is fixed at $\mu_0 H_x=-0.2$~T: a value identical to that used in the transport measurement shown in Fig.~\ref{fig:raw}.

We start the analysis by concentrating first on panel (a) of Fig.~\ref{fig:bls}, which shows in a density plot the actual variation of the magnon spectra at Pos1. The spectral variation of the BLS signal at $I_1=0$ leads experimentally to a peak instead of the predicted step function shown in Fig.~\ref{fig:be}(a) for 2D systems. The attenuation of the spectral sensitivity at high frequencies is an experimental artifact related to the extreme focus of the optical beam, which renders the scattered light insensitive to spin-waves whose wavevectors are larger than the inverse of the beam waist. The decrease of the signal above the Kittel frequency is thus directly related to the transfer function of the detection scheme, which attenuates short wavelength magnons \cite{Demidov2012,Demokritov2008,sandercock1979light}.

The black dashed lines in the plots Fig.~\ref{fig:bls}(a) show the shift of the Kittel frequency as a function of $I_1$. At low current, the shift follows a parabolic behavior as expected for Joule heating. Note that the curvature of the parabola increases as one moves away from the heat source, as shown in Fig.~\ref{fig:bls}(c) taken at Pos2, which we associate with the lateral thermal gradient $\partial_x T_1 <0$\cite{an2021short,Shan2016}. It is also worth noting that for the signal at Pos1 in Fig.~\ref{fig:bls}(a), the local curvature increases dramatically as one approaches $I_\text{c}$, which we attribute to the decrease in $M_1$ as $T_1$ approaches $T_c$.  Interestingly, the effect is more pronounced at $I_1 > 0$ than at $I_1 <0$, suggesting a self-digging effect due to asymmetric excitation of low-energy magnons. Also visible in Fig.~\ref{fig:bls}(a) is the extinction of the BLS intensity in the region \ding{194} when $I_1 \ge 2.1$~mA. Finally, the BLS signal decays at large currents at Pos1 (below the emitter), while remaining finite at Pos2 (below the collector). 

The density plot in Fig.~\ref{fig:bls}(c) further confirms the enhancement / attenuation of the spin fluctuations depending on the polarity of the current. Starting from thermal fluctuations at $I_1=0$, the signal decreases for $I_1 \cdot H_x >0$ and increases for $I_1 \cdot H_x < 0$.  A more detailed analysis at low current amplitude (not shown) confirms a linear variation in $I_1$ in the region \ding{192}. BLS spectra at large currents are shown in Fig.~\ref{fig:bls}(b). They compare the magnon distribution observed in Fig.~\ref{fig:bls}(a) at $I_1=2$~mA for both negative (green horizontal dash-dotted cut) and positive (orange horizontal dash-dotted cut) polarity of the current. At $I_1=-2$~mA, the maximum amplitude in Kittel mode (green vertical marker at 6.7~GHz) has a normalized amplitude below 1, i.e., a lower amplitude than at $I_1=0$. At $I_1=+2$~mA the density plot shows a significant enhancement of the signal, highlighted by the dashed circles in Fig.~\ref{fig:bls}(a). The corresponding amplitude of the signal [orange dots in panel (b)] shows an enhancement of more than an order of magnitude. This enhancement corresponds to the increase in spin conduction in the region \ding{193}. In addition, the BLS data taken at Pos1 show the self-localization of the asymmetric excitation due to the red shift that arises for large cone angles\cite{demidov2017chemical,divinskiy2019controlled}. It is important to note that the induced shift is almost as large as it can be, since the Kittel frequency almost reaches the paramagnetic limit $\omega_H = \gamma H_0$, visible on the panels (a) and (c) as a light vertical doted line at about 5.8~GHz \footnote{The resonance line at 5.72 GHz in Fig.~\ref{fig:bls}(a,c) probably corresponds to the paramagnetic response of the substrate. For clarity it has been removed from the plot by background subtraction.}. Note that the paramagnetic limit is reached only for $I_1 > 0$ and not for $I_1 <0$. This suggests that the film is still in its ferromagnetic phase when the signal disappears at $I_c\approx 2.1$~mA. This feature will be discussed in connection with the discrepancy between $T_c^\star$ and $T_c$ in part II. 
The excitation pocket created under the emitter within the circular area also appears as a peak at Pos2 in Fig.~\ref{fig:bls}(d), centered around the orange vertical marker at 6.7~GHz, i.e., well below the position of the Kittel mode at that position (green vertical marker at 7.4~GHz). This suggests that the collector could still probe magnetic fluctuations that are spectrally below (about 0.7~GHz) $E_g$, the energy bandgap at this position. Although the amplitude of the peak at the orange marker is significantly reduced as one moves from Pos1 to Pos2, confirming numerous indications that the majority of low-energy magnons remain localized below the emitter electrode (see below), these spectral fluctuations are not completely suppressed. We interpret this apparent contradiction as a signature that the wavelength of the mode excited under the emitter ($\lambda$ around 0.6$\mu$m) remains large compared to the width of the emitter well, and the ratio corresponds to the evanescent decay between Pos1 and Pos2. The issue of spatial decay will be investigated in more detail in part II\cite{kohno_2F}. 

To gain further insight, we plot the spectral integration of the BLS signal as a function of $I_1$ in Fig.~\ref{fig:bls}(e). The first key feature directly observed in the plot is the continuity of the signal across the origin for both polarity of the current $I_1$. This observation is used above to extract the asymmetry of the magnon transmission ratio. It will be shown below that the BLS measurement confirms some other key features observed in the magnon transport properties shown in Fig.~\ref{fig:raw}. The second key feature is the limited growth of the intensity at $I_\text{pk}$. However, the size of the peak is larger at Pos1 than at Pos2, suggesting the importance of localized magnons under the emitter. The third key feature is the idea that there are 3 transport regimes. In particular, we observe the collapse of the BLS signal at $|I_1| > |I_c|$ under the emitter.  

To emphasize the similarity with transport, in Fig.~\ref{fig:bls}(e) we tentatively plot the evolution with $I_1$ of the number of thermally activated low-energy magnons produced by Joule heating with a dashed line. The behavior follows the curve $\Delta n_{T_1}$ introduced in part II \cite{kohno_2F}. The underlying parabolic increase of this background signal is directly visible in the evolution of the maximum BLS intensity along the dashed line in Fig.~\ref{fig:bls}(e).  The blue and red shaded areas show the deviation of the BLS intensity from the nominal thermal occupancy. They indicate the amount of magnons emitted and absorbed by the STE in forward and reverse bias, respectively, and the deviation is analogous to the blue and red shaded areas in Fig.~\ref{fig:raw}(b). For $I_1 < I_c/2$ the curves deviate equally from the dashed curve, indicating that equal amounts of magnons are transferred between the metal electrode and the YIG film in the linear regime between forward and reverse bias. The curvature at the origin should scale as the variation of $M_1$ with $I_1$. To this end, we plot the evolution of $M_1$ under the emitter in the inset Fig.~\ref{fig:bls}(f). The blue points are derived from the variation with current of the spectral position of the Kittel frequency, which is sensitive to temperature. The observed behavior is consistent with the expected evolution of $M_1$ with $I_1$, which is derived from the dependence of $T_1$ on $I_1$ as shown in Fig.~S1 of Ref.~\cite{kohno_2F}. The result is shown as a dashed line and the agreement ensures the validity of the evolution with $I_1$ of the thermal background shown in Fig.~\ref{fig:bls}(e). In this data treatment, we do not attempt to correct for the use of different pulse duty cycles, which results in a small discrepancy in the value of $I_\text{c}$.

To conclude this section, we interpret the fact that all the distinct transport signatures observed in Fig.~\ref{fig:raw}(b) are present in the BLS intensity shown in Fig.~\ref{fig:bls}(e) as a strong indication that they are predominantly driven by the change in density of low-energy magnons below the emitter. However, we observe a difference in the scaling of the effect, which we attribute to the self-localization of these low-energy magnons. This will be the subject of the band mismatch below. A more rigorous quantification of the localization fraction is part of the analysis in part II.

\begin{figure}
    \includegraphics[width=0.49\textwidth]{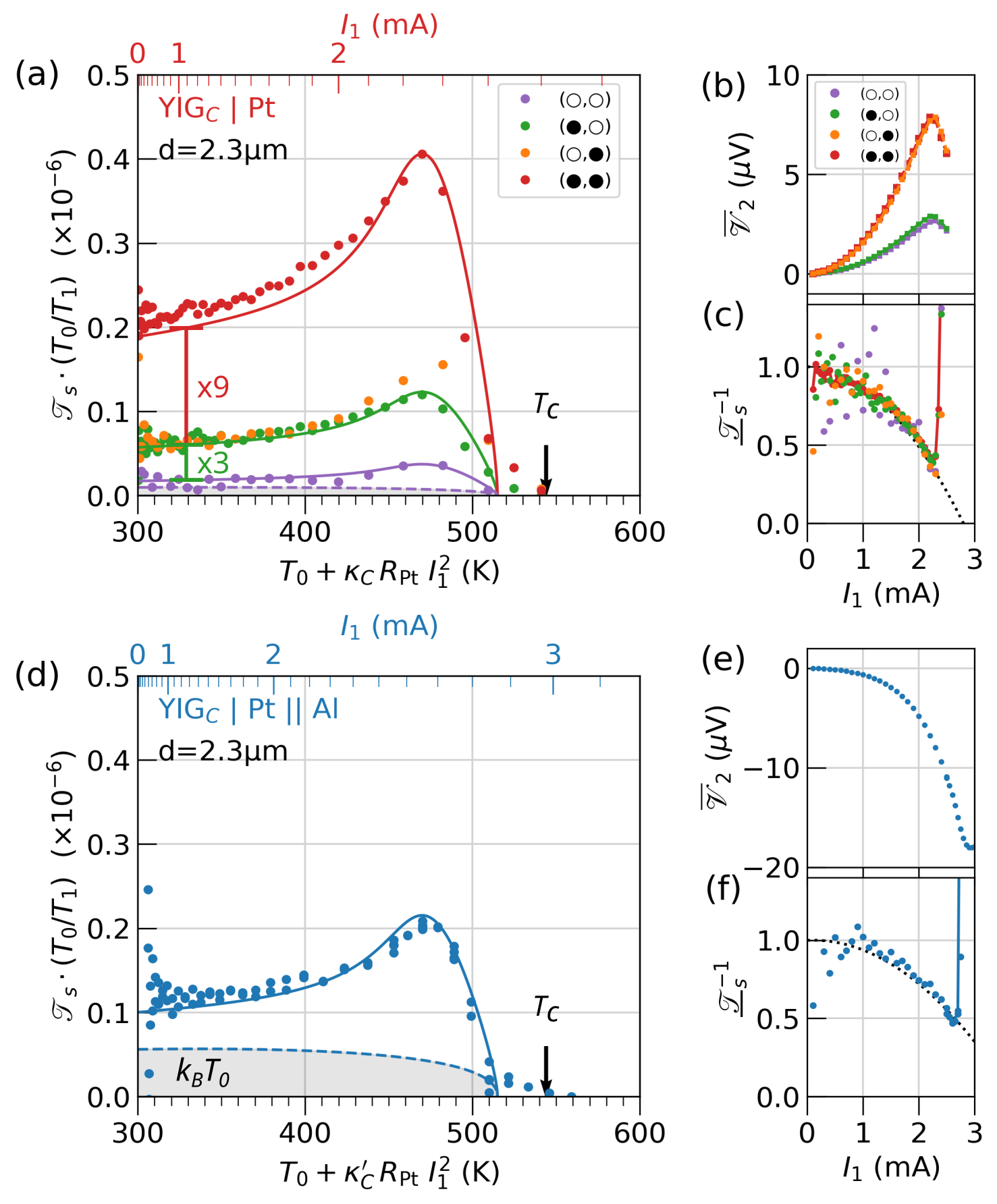} 
    \caption{Experimental evidence that the spin diode effect is mainly controlled by thermal fluctuations below the emitter. Spin diode effect as a function of emitter temperature raised by Joule heating, $T_1=T_0+\kappa_C R_\text{Pt}I_1^2$, when either (a) $\epsilon$, the spin-transfer efficiency, or (d) $\kappa_C$, the heat dissipation coefficient, are independently varied. The upper scale shows the corresponding current bias, $I_1$. The legend of panel (a) shows with the symbols $\circ$/{\textbullet} which electrode of the pair (emitter, collector) has been subjected to local annealing \cite{kohno2021enhancement} resulting in $\epsilon_\text{\textbullet}$/$\epsilon_{\circ}\approx 3$ as shown in Table.~\ref{tab:mat}. Panel (d) shows the behavior when the annealed device is covered by an Al heat sink, which reduces the efficiency of the Joule heating $\kappa_C^\prime < \kappa_C$. In panels (a) and (d), all solid lines have the same shape as in Fig.~\ref{fig:raw}(b), albeit with different scaling factors. The right panel shows the corresponding current dependence of the SSE voltage $\Vbckgnd$ in panels (b) and (e) and the normalized inverse transmission coefficient of the spin current $\mathscr{T}_s^{-1}=I_1/I_2$ in panels (c) and (f). The dashed lines are parabolic fits.}
    \label{fig:lsv4}
\end{figure}

\subsection{Influence of the lattice temperature.} \label{sec:t1}

In this section we investigate the relevant parameters that influence the spin diode effect. To this end, we propose in Fig.~\ref{fig:lsv4} to compare the asymmetric regime while independently varying either $\epsilon$, the interfacial efficiency of spin-to-charge conversion, or $\kappa$, the efficiency of device thermalization. It builds on our two recent studies \cite{an2021short} and \cite{kohno2021enhancement}, where more details can be found. The device was patterned on YIG$_C$, which is the same film as the one studied in Fig.~\ref{fig:raw}, but twice as thick as the YIG$_A$ film studied in Fig.~\ref{fig:bls} (see Table.~\ref{tab:mat}). In this batch, local annealing \cite{kohno2021enhancement} can be used to apparently increase the spin mixing conductance $G_{\uparrow\!\downarrow}$ of a single electrode. The term spin mixing conductance should be interpreted here as an effective fitting quantity. As explained by Kohno \textit{et al.}\cite{kohno2021enhancement}, the physics at play in this annealing mechanism does not involve a change in intrinsic properties, but rather a change in extrinsic properties via an expansion of the contact area. We denote $\epsilon_{\circ}$ and $\epsilon_\text{\textbullet}$ as the STE efficiency before and after annealing. We have shown that applying local annealing for 60~mins at 560~K can reduce $\epsilon_\text{\textbullet}/\epsilon_{\circ} \approx 3$ without changing the Pt resistance, i.e., $\kappa_C$ \cite{kohno2021enhancement}.  

We first investigate changes in the spin transport when alternatively the collector and emitter are annealed using the 4 possible combinatorial configurations possible as summarized in the legend of Fig.~\ref{fig:lsv4}(a).  We find in Fig.~\ref{fig:lsv4}(c) that the normalized current ratio 
\begin{equation}
    {\underline{\mathscr{T}}_s}^{-1} \equiv \left (\mathscr{T}_s \; \middle/ \; \mathscr{T}_s|_{{I_1} \rightarrow 0} \right )^{-1}
\end{equation}
of the 4 curves fall on the same parabola. Note that we have introduced the notation of underlined symbols, which will be referred to in the following as the quantity normalized to the low current value. The parabolic shape is consistent with the behavior reported in Fig.~7(c) of part II\cite{kohno_2F}, which suggests a decrease as $1-(I_1/I_c)^2$ in the \ding{193} regime. This already supports that the value of $\epsilon$ plays very little role in the relative amount of electrically excited low-energy magnons.  If the sample were well thermalized $T_1 \approx T_0$, the zero point of the parabolic fit shifts significantly to higher current values, as shown below in Fig. ~\ref{fig:lsv4}(f), suggesting that in this case the extrapolated decay ${\underline{\mathscr{T}}_s}^{-1}$ points to the amplitude of $\mathcal{I}_\text{th}$ as suggested by the inversion of Eq.~(\ref{eq:dnk}). The intercept value decreases as the influence of Joule heating, or $\kappa_C$, increases and is bounded by $I_\text{c}$ as discussed in Fig.~\ref{fig:analytic}. We recall that local annealing has not changed the Pt resistance and thus in Fig.~\ref{fig:lsv4}(a-c) the efficiency of Joule heating $\kappa_C$ is identical for all 4 cases. These conclusions are also consistent with the behavior of the SSE signal $\Vbckgnd$ shown in Fig.~\ref{fig:lsv4}(b). The 4 data sets are exactly divided into two pairs of curves, which are scaled by the ratio $(\epsilon_\text{\textbullet} / \epsilon_{\circ})$. The difference is solely determined by the value of the efficiency coefficient $\epsilon_2$ on the collector side. As expected for SSE, changes in the value of $\epsilon_1$ on the emitter side are irrelevant. 

We now plot in Fig.~\ref{fig:lsv4}(a) the variation of the renormalized STE transmission coefficient $\mathscr{T}_s \cdot T_0/T_1$ as a function of $T_1=\kappa_C R_\text{Pt} I_1^2 + T_0$, with $T_0=300$~K. The normalization by $T_1$ allows to view the magnon density changes from the point of view of a well thermalized device, as suggested by Eq. (\ref{eq:dim}). The upper abscissa scale can be used as an abacus to convert $T_1$ back to $I_1$.  On all devices we find that $\mathscr{T}_s \cdot T_0/T_1$ increases nonlinearly above 360~K. Above 460~K the spin transmission decreases with increasing temperature, as explained in Fig.~\ref{fig:raw}(b). 
As expected, we find that the amplitude of $\mathscr{T}_s \cdot T_0/T_1$ scales as the product of $\epsilon_1 \cdot \epsilon_2$. This observation confirms that the Pt is weakly coupled to the YIG, i.e., only a small part of the spin current circulating in the YIG is detected by the collector. If this were not the case, the signal would not depend on the collector efficiency $\epsilon_2$. This is further confirmed by the width dependence of $\mathscr{T}_s$ (see Fig.~\ref{fig:widthdep}). 

The most interesting feature is the approximate superposition of the 2 curves ($\circ$,\textbullet) and (\textbullet,$\circ$), which consists in inverting emitter and collector in a device where one electrode is 3 times more efficient at emitting magnons. While the superposition is almost perfect at low currents, at high currents the magnon conductance with emitter and collector inverted (orange dots) leads to a significantly larger magnon conductance than the normal configuration (green dots). However, the difference is small compared to the factor of 3 produced by annealing. Thus for all practical purposes, we interpret the data as somewhat confirming that the spin current circulating between the contact electrodes is proportional to $\epsilon_1 \cdot \epsilon_2$, as Eq.~(\ref{eq:dim}) explains it. The superposition of the two curves in Fig.~\ref{fig:lsv4}(a) shows that the density of magnons in the YIG (here it is varied by a factor of 3) seems to have no effect on the shape of $\mathscr{T}_s \cdot T_0/T_1$, and the pertinent parameter that determines its behavior remains the emitter temperature $T_1$. Moreover, if one compares in Fig.~8 of \cite{kohno_2F} $\mathscr{T}_s$ versus $T_1$ (as opposed to as versus $I_1$) taken on two different YIG thicknesses, it seems that the thickness plays no role in determining the nonlinear behavior between $I_2$ and $I_1^2$. 

To further support this notion, the coefficient $\kappa_C$ can be modified while keeping $\epsilon$ constant. As explained in \cite{an2021short}, one can cover the annealed device with a 105~nm thick Al layer that acts as a heat sink (the heat sink is separated from the Pt electrode by a 20~nm thick protective layer of Si$_3$N$_4$). The influence of the heat sink can be seen by comparing the abacus in Fig.~\ref{fig:lsv4}(a) and Fig.~\ref{fig:lsv4}(d). The Al layer has reduced the efficiency of the Joule heating $\kappa_C$ by 27 $\%$. The changes are most visible in Fig.~\ref{fig:lsv4}(e), which shows a completely different nature of the SSE signal compared to Fig.~\ref{fig:lsv4}(b) (different curvature, different polarity \cite{an2021short}). However, once $I_1$ is converted to $T_1$, Fig.~\ref{fig:lsv4}(d) scales with Fig.~\ref{fig:lsv4}(a), even though a larger amount of current $I_1$ is circulating in the former. 

This leads to the conclusion that the nonlinearity of the magnon transmission ratio $\mathscr{T}_s \cdot T_0/T_1$ produced by STE seems to be governed mainly by the lattice temperature. This observation alone seems to contradict our previous conclusion from the BLS experiment, which attributes the spin diode effect to low-energy magnons. The apparent discrepancy is explained in Fig.~\ref{fig:analytic}(c). Knowing that the density of low-energy magnons is also sensitive to temperature, the disappearance of $M_1$ at $I_\text{c}$ due to Joule heating shifts the position of $I_\text{pk}$ below $I_\text{c}$, which then becomes weakly dependent on the room temperature value of the threshold current, $\mathcal{I}_{\text{th},0}$. The process still requires $\mu_M$ to approach $E_g$ by injecting spins, but once in the vicinity and the nonlinear process is fully in place, the temperature change is dominant. Thus, this artifact is a consequence of the inability to efficiently thermalize the Pt electrode rather than an intrinsic phenomenon. Nevertheless, it emphasizes the importance of thermal fluctuations in nonlinear phenomena. It underlines that the parametric threshold is also determined by the initial magnon thermal fluctuation in the sample, bearing in mind that all nonlinearities are suppressed at absolute zero temperature.

All experimental data in Fig.~\ref{fig:lsv4}(a) and Fig.~\ref{fig:lsv4}(d) are fitted with the same curve as shown in Fig.~\ref{fig:raw}(b). The only parameter that varies is the vertical scaling factor. 
The fact that all the curves have exactly the same shape also supports the suggestion that $T_1$ determines the nature of $\mathscr{T}_s \cdot T_0/T_1$, where the shaded region shows the background contribution from high-energy thermal magnons $\mathscr{T}_T$, where $\Sigma_T/(\Sigma_T+\Sigma_K)\approx 0.5$ represents the relative weight at this distance. 

The Al capping also changes the spatial profile of $T(\mathbf{r})$, which defines the confinement potential of $M_T$ near the emitter, as shown in Ref.~\cite{an2021short}. Thus, the absence of a large discrepancy between Fig.~\ref{fig:lsv4}(a) and Fig.~\ref{fig:lsv4}(d) indicates that the depth of band shift produced by Joule heating is the dominant parameter, while the spatial extent of the confinement region defined here by $\partial_x T$ is not significant.

\subsection{Magnon band mismatch.}

In this section we will further elucidate which nonlinear effects lead to the suppression of spin propagation in the high power regime.

\begin{figure}
    \includegraphics[width=0.49\textwidth]{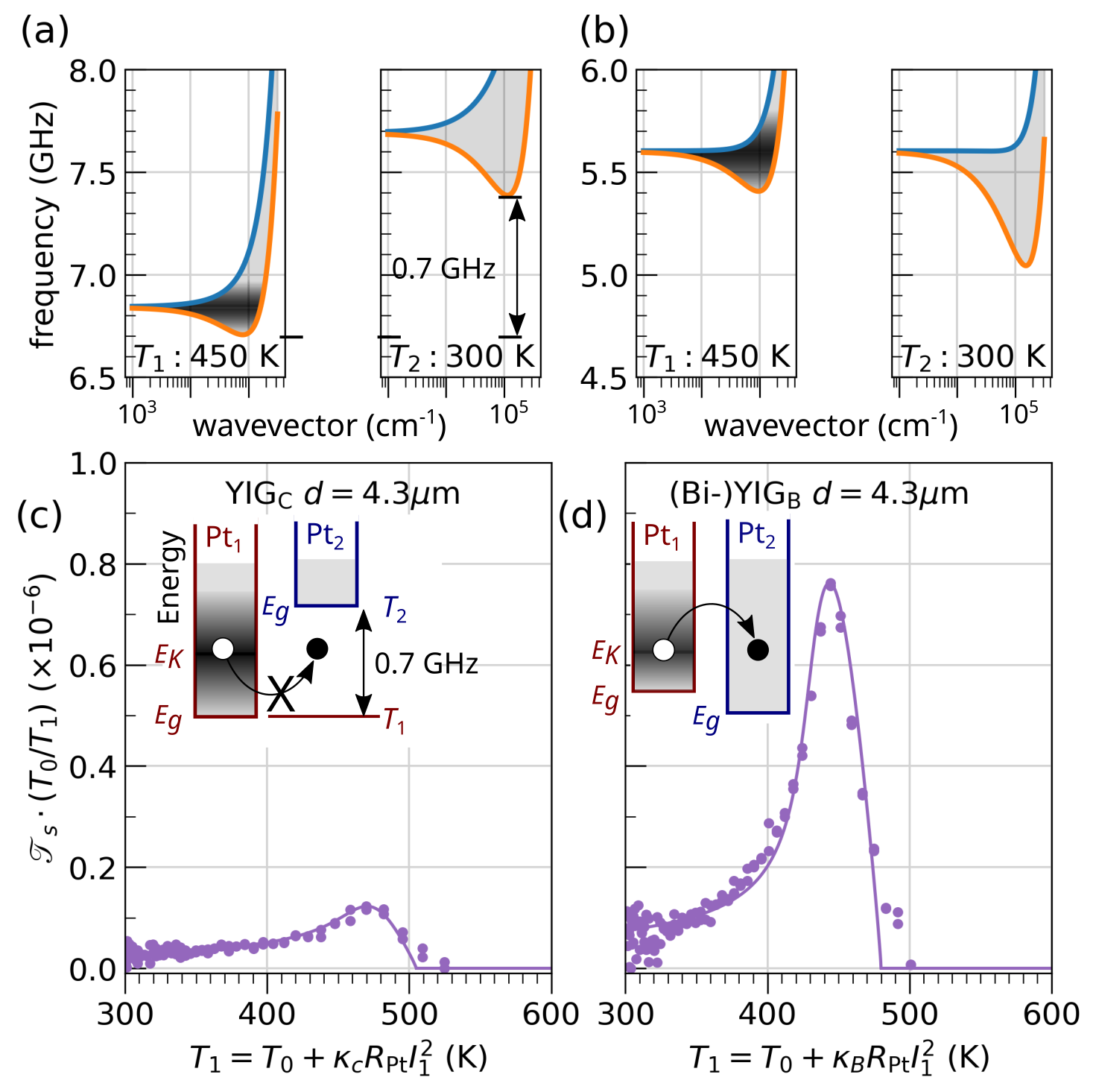}
    \caption{Experimental evidence that the spin diode effect involves a narrow spectral window of magnons around the Kittel energy. The upper panels illustrate the band mismatch in the GHz spectral range between two regions at different temperatures. In panel (a), calculated for YIG$_C$ thin films, a propagation gap of almost 0.7~GHz rises between the low-lying magnon modes, when the temperature difference attains 150~K. There, the frequencies of the low-energy magnons below the emitter are below the bulk bandgap and thus prevented from propagating and instead encouraged to self-localize, limiting the growth of the nonlocal signal. In panel (b), calculated for (Bi)-YIG$_B$, this barrier vanishes when $K_u - \mu_0 M_s \rightarrow 0$. There, the low-energy magnons are allowed to contribute significantly to the long-range spin transport. Comparison of the propensity of low-energy magnons to travel long distances ($d=4.3$ $\mu$m) between (c) YIG$_C$ and (d) (Bi-)YIG$_B$. We observe a significant increase in the spin diode signal between (c) YIG$_C$ and (d) (Bi-)YIG$_B$. We explain this by the suppression of self-localization effects produced by the nonlinear frequency shift.}
    \label{fig:nlfreq}
\end{figure}

\subsubsection{Nonlinear frequency shift.}

We begin this discussion by examining the issue of nonlinear frequency shift. We recall that this effect comes from the depolarization factor along the film thickness $N_{zz} = -1$, which introduces a correction to the Kittel frequency that depends on the magnetization, $M_s$. This coefficient corresponds to a red shift, i.e., a decrease of $M_s$ causes a red shift of the Kittel mode and thus of $E_g$. The maximum shift that can be induced is $\left ( \omega_K - \gamma H_0\right )/(2\pi) \approx 2$~GHz, which is the energy difference between the Kittel mode and the paramagnetic limit at $\mu_0 H_0 = 0.2$~T. Fig.~\ref{fig:bls} shows experimental evidence for shifts as large as 0.7~GHz, produced either by Joule heating of the emitter [see the dashed line in Fig.~\ref{fig:bls}(a) and (c)] or by self-localization of the mode (see the dashed circle in Fig.~\ref{fig:bls}). An increase in the precession angle $\theta$ also produces a red shift of the resonance due to a decrease in the internal field, which depends on the time-averaged magnetization following $M_s \cos \theta$\cite{Bauer2015,barsukov2019giant}. The latter effect is responsible for the foldover of the main resonance. As shown in the inset of Fig.~\ref{fig:nlfreq}(c), such a shift is strong enough to push the entire spectral window of low-energy magnons fluctuating below the emitter underneath the energy gap $E_g$ of the YIG film outside. This suggests that spin fluctuations produced by STE can still be sensed below the collector despite having energies below the propagation band window of magnons at this position, and the propagation length $\lambda_K$ is the decay length of the evanescent spin-wave outside the energy well. This favors the self-localization effect\cite{Slavin2005b,ulrichs2020chaotic,schneider2021stabilization} as reported in the analysis of Fig.~8 of Ref.~\cite{kohno_2F}.

We want to compare this behavior with the transport properties in (Bi-)YIG$_B$ thin films (see material parameters in Table.~\ref{tab:mat}). The peculiarity of sample B is that the YIG sample is doped with Bi.  For the right concentration of Bi it is possible to match the uniaxial anisotropy with the saturation magnetization (see Table.~\ref{tab:mat}), which leads to $M_\text{eff} \approx 0$. This corresponds to a thin film that has an isotropically compensated demagnetization effect. In this case, the Kittel mode becomes isochronous and independent of the precession cone angle. The Kittel frequency simply follows the paramagnetic proportionality relation $\omega_K = \gamma H_0$ (similar to the response of a sphere). In particular, the value of $\omega_K$ is independent of $M_T$ and therefore the nonlinear frequency shift is null\cite{divinskiy2019controlled,Evelt2018}. This implies that the conduction band of the magnons between emitter and collector remains aligned as schematically shown in Fig.~\ref{fig:nlfreq}(d), which then prevents the self-localization effect of the nonlinear frequency shift. The improvement of the magnon transmission ratio is clearly visible when comparing Fig.~\ref{fig:nlfreq}(c) and Fig.~\ref{fig:nlfreq}(d). Further evidence for the reduction of self-localization in the latter case is the observation that the variation of the transmission ratio in Fig.~\ref{fig:nlfreq}(d) now mimics the observed variation of low-energy magnons under the emitter by BLS, as shown in Fig.~\ref{fig:bls}(e). We will return to this important point in PartII\cite{kohno_2F} while discussing the spatial decay of the magnon transmission ratio.

On the quantitative side, we find that the ratio of initial to maximum values is 15.1 for (Bi-)YIG$_B$ compared to 7.2 for YIG$_C$). A first important observation is that while the suppression of the elliptical precession and the temperature dependence of $\omega_K$ by the uniaxial anisotropy compensating the dipolar field favors the population of low-energy magnons, the signal still saturates in the case of (Bi-)YIG$_B$.  The solid line in Fig.~\ref{fig:nlfreq}(c) and (d) is a fit with Eq.~(\ref{eq:dnk}) using $n_\text{sat}$ as a free parameter. While for the YIG$_C$ sample we find that the best fit is obtained by using $n_\text{sat}=4$ as explained above, the value of the saturation threshold increases to $n_\text{sat}=11$ in the case of Bi-YIG$_B$. This is also a direct experimental evidence that the contribution of low-energy magnons to the spin diode effect concerns a rather narrow spectral window around the Kittel energy, with an upper limit width of about 1~GHz. This broadening should be correlated with the enhanced magnon-magnon scattering time indicated by $\Gamma_m$ in Fig.~\ref{fig:be}(c). Another relevant energy scale is the difference $E_K - E_g$, which varies as $t_\text{Y1G}$. This implies that in ultrathin films of garnet this phenomenon of localization of low-energy magnons is enhanced.

\subsubsection{Saturation of the magnon density.}

We emphasize, however, that although the nonlinear frequency shift is zero in the case of (Bi-)YIG$_B$, the system is still subject to the saturation effect. For example the observed factor 15.1 is still significantly smaller than the change in cone angle observed in nanopillars at the damping compensation threshold. This implies that the aforementioned saturation effects and the rapid growth of the magnon-magnon relaxation rate is a general phenomenon that is not suppressed by reducing the nonlinear frequency shift. The vanishing nonlinear frequency coefficient only eliminates the ellipticity of the trajectory for the long wavelength magnons whose wavevector is smaller than the inverse of the film thickness. It does not eliminate the self-depolarization effect of the spin-wave. This value depends mainly on $\theta_k$, the angle between the propagation direction and the equilibrium magnetization direction, the latter being the origin of the magnon manifold broadening. We note that for the wavevectors around $E_g$ this is the dominant origin of the ellipticity, since the broadening almost reaches the maximum value of $\gamma \mu_0 M_s$, as shown in Fig.~\ref{fig:intro}(d).
 
 This shows that the tuning of $M_\text{eff}$, which allows to remove the non-isochronicity of the long wavelength magnons, is responsible for an increase of the saturation threshold, which is found to be almost three times higher. 

\section{Conclusion.}

In this work we draw a comprehensive picture of the role of low-energy magnons in the electrical transconductivity of extended magnetic thin films. While spin conduction at low intensities appears to behave largely as expected, the behavior at high intensities is markedly different from that observed in highly confined geometries such as nanopillars. 
 
The main difference is related to the tendency of the injected spin to spread between different degenerate eigenmodes. Thus, there are phenomena related to the two-dimensionality that intrinsically prevent a single mode from dominating the others (as is possible in 0D and 1D \cite{divinskiy2021evidence}). In a confined geometry, the energy gap created by confinement between different eigenmodes protects the main fluctuator from relaxation into other modes. In an extended thin film, this barrier is removed and degeneracies arise, leading to an efficient redistribution of energy between degenerate modes. Even if one mode is pumped more efficiently than the others (e.g. the mode with wavelength $1/w_1$), nonlinear saturation phenomena quickly come into play, so that the critical current can never be reached. This leads to the magnon-magnon relaxation rate becoming power dependent, and in particular to a sharp increase above a certain mode occupation threshold. We use this to paint a picture of a condensate that appears to behave like a liquid. This picture is supported by a number of different experiments, including nonlocal transport on different thicknesses of YIG thin films as well as different garnet compositions, Brillouin light spectroscopy, independent variation of the spin mixing conductance or the thermal gradient near the emitter.

We have shown that doping with Bi improves the nonlocal signal. The first reason is that we avoid the nonlinear red shift under the emitter, which produces localization, which is obviously detrimental to the nonlocal geometry. The second reason is that the lateral temperature gradient has no effect on the magnon spectrum, since $M_\text{eff} \approx 0$ regardless of the temperature. So the magnons excited under the emittor have no problem propagating to the collector. And the third reason is that since the precession is quasi-circular, the parametric excitation of other modes of magnons is strongly limited, allowing $n_\text{sat}$ to become larger.

We have also shown that the inability to thermalize the emitter electrode plays a crucial role in the decrease of $M_1$ under the emitter. This is a very strong effect in 2D geometry which, combined with the fact that a single mode cannot be excited and the critical current goes to infinity, means that we reach the Curie temperature before it self-oscillates. Although this reduction in $M_1$ may seem favorable for reaching the critical current (which tends to 0 as $M_1 \rightarrow 0$), the nonlinear effects mentioned above (coupling between modes, location under the emitter, etc.) make it inaccessible.

We have not found any direct signature of BEC in our transport studies on nonlocal devices. All of our experimental data point to strong interaction between degenerate modes rather than fluctuation of a single mode. This problem plagues the nonlocal geometry, where one only observes the magnon propagating outside the Pt electrode, which, represents only a small fraction of the total injected spin. In this respect, our work above does not provide answers about what happens directly under the emitter.

Although there is no BEC condensation outside the area below the emitter, the analogy with the Gurzhi effect in an ultrapure electron conductor seems appropriate. Further studies would be required to provide direct evidence for such magneto-hydrodynamic fluid behavior at high power. The unique signature would be features that can only be attributed to the Navier-Stokes transport equation \cite{Polini2020}.

\begin{acknowledgments}
This work was partially supported by the French Grants ANR-18-CE24-0021 Maestro and ANR-21-CE24-0031 Harmony;  the EU-project H2020-2020-FETOPEN k-NET-899646; the EU-project HORIZON-EIC-2021-PATHFINDEROPEN PALANTIRI-101046630. K.A. acknowledges support from the National Research Foundation of Korea (NRF) grant (No. 2021R1C1C201226911) funded by the Korean government (MSIT). This work was also supported in part by the Deutsche Forschungs Gemeinschaft (Project number 416727653).

\end{acknowledgments}

\newpage

\section{Annex}

\setcounter{figure}{0}
\renewcommand{\thefigure}{S\arabic{figure}}%
\renewcommand{\thetable}{S\arabic{table}}%

\subsection{Sample characterization}

All of the magnetic garnet films used in this study have their macroscopic magnetic properties fully characterized. Curves of magnetization versus temperature are shown in the appendix of \cite{kohno_2F}. The same is true for the Pt metal electrode, whose resistivity and its dependence on temperature are given in the same appendix.  All values are summarized in Table.~\ref{tab:mat}. 

\begin{figure}
    \includegraphics[width=0.49\textwidth]{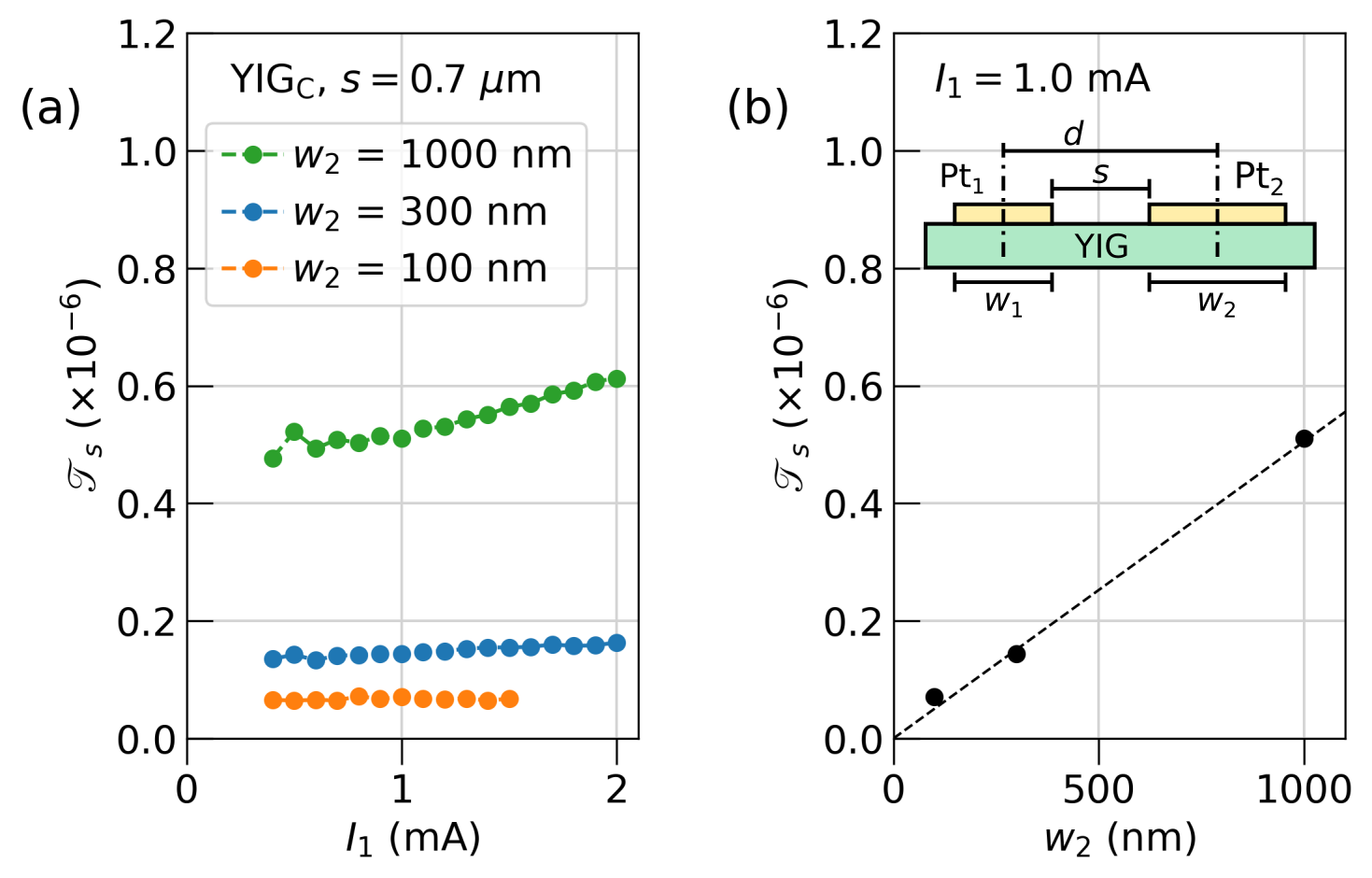}
    \caption{Impact of collector width, $w_2$. The change of $w_2$ occurs while keeping both the emitter width, $w_1 = 300$~nm, and the edge to edge distance between the two Pt strips, $s=d-(w_1+w_2)/2=0.7$~$\mu$m, where $d$ is the center to center distance. (a) shows $\mathscr{T}_s$ as a function of $I_1$ for 3 different values of $w_2$ and (b) shows $\mathscr{T}_s$ at $I_1=1$ mA as a function of $w_2$. The dashed line is a linear fit through the data point. The inset is a schematic side view of the devices, defining the various dimensions used throughout the paper.}
    \label{fig:widthdep}
\end{figure}

\begin{figure}
    \includegraphics[width=0.49\textwidth]{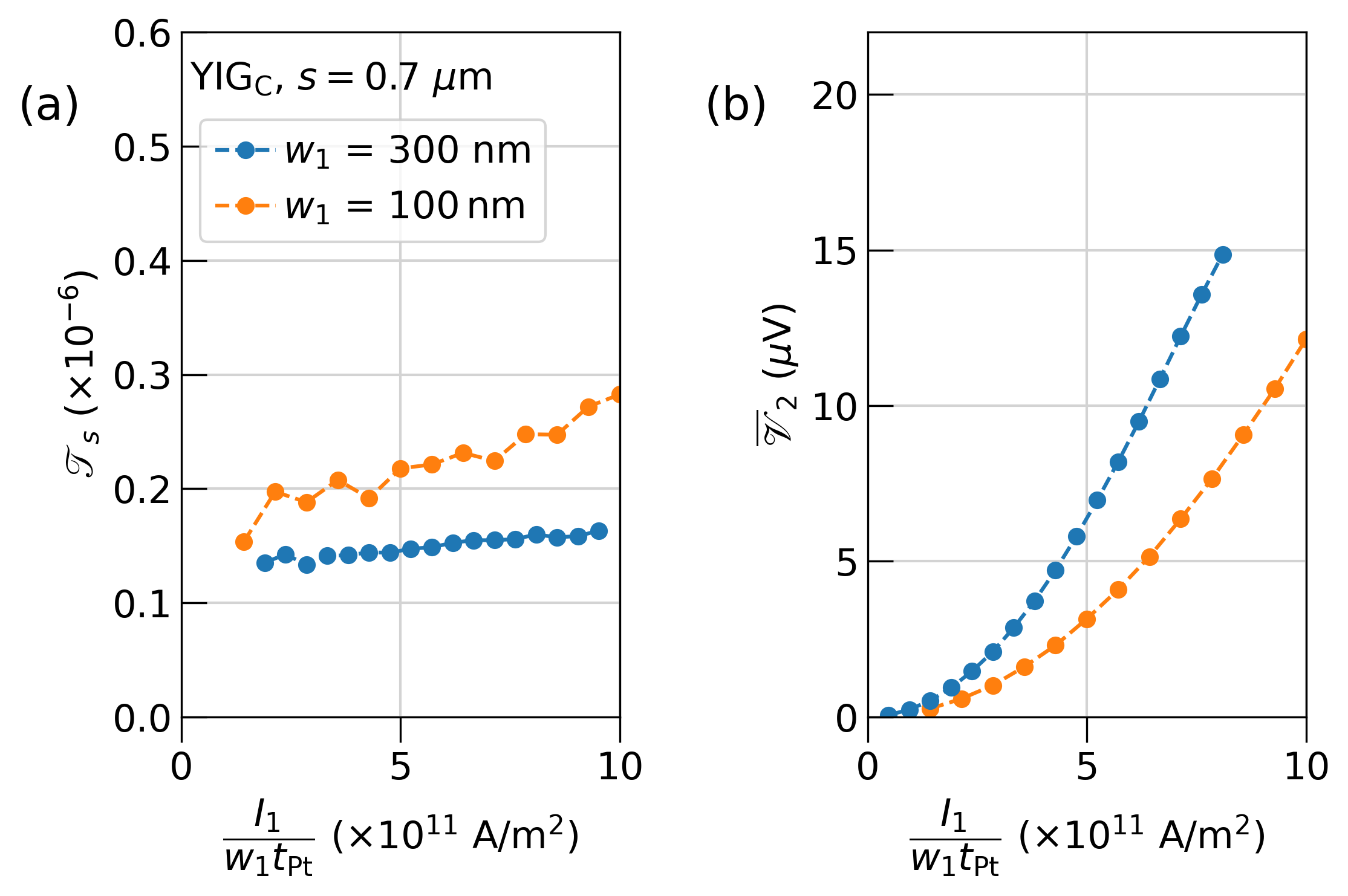}
    \caption{Impact of emitter width, $w_1$. The change of $w_1$ occurs while keeping both the collector width, $w_2 = 300$~nm, and the edge to edge distance between the two Pt strips, $s=d-(w_1+w_2)/2=0.7$~$\mu$m, constant. (a) shows $\mathscr{T}_s$ as a function of the current density $J_1=I_1/( w_1 t_{\mathrm{Pt}})$ for 2 different values of $w_1$ and (b) shows the $\Vbckgnd$ voltage ($\propto$ SSE) as a function of $T_1$.}
    \label{fig:injwidthdep}
\end{figure}

Charge to spin (or vice versa) interconversion is provided by the Spin Hall effect. Its efficiency process is described by the relation 
\begin{equation}
\epsilon \equiv \dfrac{G_{\uparrow\!\downarrow} \,\theta_\text{SHE}\, \tanh \left [ t_\text{Pt} /({2 \lambda_\text{Pt}}) \right ] }{ G_{\uparrow\!\downarrow} \coth \left ( t_\text{Pt} / {\lambda_\text{Pt}} \right ) + {\sigma_\text{Pt}} / (G_0 \lambda_\text{Pt}) } \hspace{0.2cm}, \label{eq:epsilon}
\end{equation}
where $\theta_\text{SHE}$ is the spin Hall angle, $G_{\uparrow\!\downarrow}$ is the spin mixing conductance, $t_\text{Pt}$ is the thickness of the Pt layer, $G_0=2 e^2/h$ is the quantum of conductance, and $\lambda_\text{Pt}$ is the internal spin diffusion length. To calculate the value of $\epsilon$, we assume that $\lambda_\text{Pt}=3.8$~nm \cite{Collet2016} and $\lambda_\text{Pt} \theta_\text{SHE}=0.18$~nm \cite{kohno2021enhancement,Sanchez2014,Sanchez2013}. From Table.~\ref{tab:mat} we see that the value of $G_{\uparrow\!\downarrow}\ll {\sigma_\text{Pt}/ (G_0 \lambda_\text{Pt}})$ and thus Eq.~(\ref{eq:epsilon}) reduces to $\epsilon \approx \theta_\text{SHE} T$ with proportionality $T= G_{\uparrow\!\downarrow} \cdot G_0 \lambda_\text{Pt} /\sigma_\text{Pt} \approx0.1 $ which is maximized when $t_\text{Pt}\approx7$~nm. Numerical evaluations of $\epsilon$ are found in Table.~\ref{tab:mat}. The maximum achieved value is $\epsilon \approx 0.08$ observed in (Bi-)YIG$_B$.

Note that for Pt, $\theta_\text{SHE} > 0$. This means that the interconversion is governed by the right-hand rule. Using the convention of Fig.~\ref{fig:intro}, a positive current (i.e., circulating along $-\hat y$) injects spins polarized along $+\hat x$, into an adjacent layer. Thus, the amplification of spin fluctuations requires that $M_s$ is aligned with $-\hat x$ or that $H_x <0$, as indicated in the figure.

\subsection{Linear regime}

Finally, it seems relevant to emphasize that the data presented above provide some cross-checking of the dimensional dependence of the magnon transmission ratio with external parameters, as suggested by Eq.~(\ref{eq:dim}). The dashed line in Fig.~7 of Ref.~\cite{kohno_2F} shows the proportional dependence of $\mathscr{T}_s$ on $T_1$. Comparison of the nonlocal signals using local annealing of the Pt electrode as shown in Fig.~\ref{fig:lsv4}(a) shows the proportional dependence of $\mathscr{T}_s$ on $\epsilon_1 \cdot \epsilon_2$. The Groeningen group has extensively studied the thickness dependence of the conductivity of electrically excited magnons (low-energy magnons) and they observed the monotonous decrease with increasing thickness, confirming the $1/t_{\mathrm{YIG}}$ behavior\cite{Shan2016}.

Eq.~(\ref{eq:dim}) also predicts the linear and inverse linear relationship of the magnon transmission ratio with the width of the collector $w_2$ and the emitter $w_1$. Fig.~\ref{fig:widthdep} shows the influence of $w_2$ on $\mathscr{T}_s$ with different $w_2$ as a function of $I_1$ for (a) and of $w_2$ for (b). Note that the width of the emitter and the edge to edge distance between the two Pt strips remain constant as $w_1=300$~nm and $s=d-(w_1+w_2)/2=0.7$~$\mu$m. These devices are fabricated at the same time as the YIG$_C$ devices and we verify that the spin mixing conductance $G_{\uparrow\!\downarrow}$ for each $w_2$ is relatively similar. Taking the value at low current $I_1=1.0$~mA to see the behavior of the linear regime, the proportional dependence on $w_2$ is revealed. This observation suggests that the YIG|Pt interface is weakly coupled, i.e., the angular momentum transfer between YIG and Pt can be considered as ineffective due to the poor transparency at the interface, which leads to only a small fraction of magnons being absorbed into the Pt electrodes. This is consistent with the observation in Fig.~\ref{fig:lsv4}(a) that the magnon emission is proportional to $\epsilon_1$. On the contrary, Fig.~\ref{fig:injwidthdep} shows the influence of the emitter width $w_1$ on (a) $\mathscr{T}_s$ as a function of the applied current density $I_1/(w_1 t_{\mathrm{Pt}})$. For the sake of completeness, we show in Fig.~\ref{fig:injwidthdep}(b) its influence on the $\Vbckgnd$ voltage as a function of the current density. The behavior shows that the thermalization of Pt$_1$ improves with decreasing emitter width. Fig.~\ref{fig:injwidthdep}(a) qualitatively confirms that the smaller $w_1$ gives greater conduction of low-energy magnons. This enhancement is due to the reduction of the effective number of spins to compensate for the damping by the spin orbit torques. If one were to convert the abscissa of Fig.~\ref{fig:injwidthdep}(b) to $T_1$, one would find that the SSE voltage does not seem to depend much on $w_1$, since the amount of temperature gradient (or Joule heating) is important. However, the fit does not scale as well as the predicted $1/w_1$. The origin of the deviation is still unclear, but one can speculate that the additional increase in magnetic damping $\alpha_\text{LLG}$ induced at the YIG$\vert$Pt interface\cite{Beaulieu2018,bertelli2021imaging} may affect the transport, especially for thermal magnons, due to its diffusive nature and short decay length of $\lambda_T\approx0.5$ $\mu$m, which is comparable to the variation range of $w_1$. Further studies are needed to clarify this point.

\bibliography{newlib}
\end{document}